%% file: main.tex
\documentclass[manuscript,screen,acmlarge]{acmart}%anonymous

\usepackage{graphicx}%
\usepackage{multirow}%
\usepackage{amsmath,amsfonts}%
\usepackage{mathrsfs}%
\usepackage{xcolor}%
\usepackage{textcomp}%
\usepackage{manyfoot}%
\usepackage{booktabs}%
\usepackage{algorithm}%
\usepackage{algorithmicx}%
\usepackage{algpseudocode}%
\usepackage{listings}%
\usepackage{hhline}
\usepackage{boldline}
\usepackage{tabularx}
\usepackage{colortbl}
\usepackage[outdir=./]{epstopdf}

\usepackage{subcaption} % 
\AtBeginDocument{%
  }

\setcopyright{acmlicensed}
\copyrightyear{2024}
\acmYear{2024}
\acmISBN{978-1-4503-XXXX-X/18/06}

\begin{document}

\title{Predicting Quality of Video Gaming  Experience Using Global-Scale Telemetry Data and Federated Learning}

\author{Zhongyang Zhang}
\email{zzy@ucsd.edu}
\orcid{1234-5678-9012}
\affiliation{%
  \institution{UC San Diego}
  \city{San Diego}
  \state{CA}
  \country{USA}
}

\author{Jinhe Wen}
\affiliation{%
  \institution{UC San Diego}
  \city{San Diego}
  \state{CA}
  \country{USA}
}
\email{jhw@ucsd.edu}

\author{Zixi Chen}
\affiliation{%
  \institution{UC San Diego}
  \city{San Diego}
  \state{CA}
  \country{USA}
}

\author{Dara Arbab}
\affiliation{%
  \institution{Arizona State University}
  \city{San Diego}
  \state{CA}
  \country{USA}
}

\author{Sruti Sahani}
\affiliation{%
  \institution{Intel}
  \city{San Diego}
  \state{CA}
  \country{USA}
}

% \author{William Lewis}
% \affiliation{%
%   \institution{Intel}
%   \city{San Diego}
%   \state{CA}

\author{Kent Giard}
\affiliation{%
  \institution{Intel}
  \city{San Diego}
  \state{CA}
  \country{USA}
}

\author{Bijan Arbab}
\affiliation{%
  \institution{Intel}
  \city{San Diego}
  \state{CA}
  \country{USA}
}

\author{Haojian Jin}
\affiliation{%
  \institution{UC San Diego}
  \city{San Diego}
  \state{CA}
  \country{USA}
}

\author{Tauhidur Rahman}
\affiliation{%
  \institution{UC San Diego}
  \city{San Diego}
  \state{CA}
  \country{USA}
}

\renewcommand{\shortauthors}{Zhongyang et al.}

\newcommand{\jinhe}[1]{\textcolor{magenta}{#1}}
\newcommand{\zzy}[1]{\textcolor{blue}{#1}}
\newcommand{\zixi}[1]{\textcolor{red}{#1}}
\newcommand{\haojian}[1]{\textcolor{orange}{#1}}
\newcommand{\sssec}[1]{\vspace*{0.05in}\noindent\textbf{#1}}
\settopmatter{printacmref=false}

\include{contents/0-abstract}

\received{20 February 2007}
\received[revised]{12 March 2009}
\received[accepted]{5 June 2009}

\maketitle

\input{contents/1-intro} 
\input{contents/2-related-work}

\input{contents/3-dataset}

\input{contents/4-data-insights}

\input{contents/5-predictor}

\input{contents/6-results}
\input{contents/7-discussion}

\input{contents/ethical}
\input{contents/8-conclusion}

\bibliographystyle{ACM-Reference-Format}
\bibliography{ref}

\end{document}

%% file: contents/0-abstract.tex
% \section{Abstract}

%DARA MODIFICATION
% \iffalse
% The Quality of a gaming experience is critically influenced by the frame rate, or frames per second (FPS), which is shaped by a complex interplay of device-specific and socio-economic factors. There are few studies done that reveal the impact of relevant factors on gaming experience. Traditional FPS prediction models are limited by their reliance on simplistic device-based calculations that fail to capture external nuances. This method can only provide rudimentary performance classifications. This paper introduces a novel predictive model with enhanced prediction accuracy. This improvement is achieved by leveraging advanced machine learning techniques such as few-shot learning and federated learning, while maintaining user privacy. This distributed FPS prediction model significantly surpasses existing methods in both accuracy and detail by utilizing a unique dataset collected from over 170 countries, 100,000 users, and nearly 1,000 games. It also offers a more personalized and privacy-aware gaming experience. 
% \fi
% there is a lack of in-depth research on the factors influencing FPS and on 
\begin{abstract}

% Frames Per Second (FPS) significantly affects the gaming experience. Letting players have an accurate FPS estimation before purchasing benefit both players and game developers. However, we have a limited understanding of how to predict a game's FPS performance on a specific device before purchase.

Frames Per Second (FPS) significantly affects the gaming experience. Providing players with accurate FPS estimates prior to purchase benefits both players and game developers. However, we have a limited understanding of how to predict a game's FPS performance on a specific device.
In this paper, we first conduct a comprehensive analysis of a wide range of factors that may affect game FPS on a global-scale dataset to identify the determinants of FPS. This includes player-side and game-side characteristics, as well as country-level socio-economic statistics. Furthermore, recognizing that accurate FPS predictions require extensive user data, which raises privacy concerns, we propose a federated learning-based model to ensure user privacy. Each player and game is assigned a unique learnable knowledge kernel that gradually extracts latent features for improved accuracy. We also introduce a novel training and prediction scheme that allows these kernels to be dynamically plug-and-play, effectively addressing cold start issues.
To train this model with minimal bias, we collected a large telemetry dataset from 224 countries and regions, 100,000 users, and 835 games. Our model achieved a mean Wasserstein distance of 0.469 between predicted and ground truth FPS distributions, outperforming all baseline methods.

\end{abstract}

\ccsdesc[500]{Computing methodologies~Neural networks}
\ccsdesc[500]{Mathematics of computing~Exploratory data analysis}
\ccsdesc[300]{Information systems~Data mining}
\ccsdesc[300]{Applied computing~Computer games}

% \ccsdesc[300]{Security and privacy~Domain-specific security and privacy architectures}
\keywords{Game Experience, Frames Per Second, Federated Learning, Deep Learning, Dataset, Video Games, Human Behavior Analysis}

%% file: contents/1-intro.tex
\section{Introduction}

Frames Per Second (FPS) significantly impacts video gaming experiences due to a complex interplay of factors~\cite{claypoolEffectsFrameRate2006, claypool2007frame, claypoolPerspectivesFrameRates2009, liuEffectsFrameRate2023}. It particularly influences player performance in competitive games, such as first-person shooters, by affecting reaction speed, distance, and accuracy~\cite{wangEffect2023}. Providing estimated FPS for a game helps customers set realistic expectations for gaming experiences on their devices, while also aiding game companies in reducing negative reviews caused by users’ misunderstandings of hardware limitations~\cite{WoLongDev,WhyCyberpunk2077}.

Currently, mainstream game platforms either provide a list of minimum and recommended configurations for manual checking~\cite{EpicGamesStore2024, SteamStore} or offer a binary prediction on whether a game will run well on a device~\cite{XboxGames}. However, these guidance are either time-consuming for users to verify or lack clear standards for interpretation.
% Currently, mainstream game platforms either provide only minimum and recommended hardware configurations for manual checking or offer a binary prediction on whether a game will run well on a device~\cite{EpicGamesStore2024, SteamStore, XboxGames}. The first approach lacks simplicity and intuitiveness, while the second has vague standards and low explainability (see Fig. \ref{fig:steam-xbox}).

% Currently, mainstream game platforms present pre-purchase configuration guidance in two ways. The first approach lists minimum and recommended parameters (e.g., graphics) for manual checking~\cite{EpicGamesStore2024, SteamStore}, which is time-consuming and requires extra knowledge. The second approach provides a simple ``game runs well'' indicator~\cite{XboxGames} for compatibility, which is more intuitive but often lacks detailed performance standards.

This work aims to provide a novel solution by studying the determinants of FPS and designing an FPS distribution predictor for players based on these findings. We thoroughly explored a range of FPS determinants, including hardware specifications, game characteristics, and socio-economic factors. Research indicates that the 95\% FPS floor is crucial for evaluating player satisfaction~\cite{liuEffectsFrameRate2023}, as it more accurately reflects the extent and frequency of lag compared to mean FPS. Our dataset analysis revealed that, at the micro level, factors such as device type, hardware specifications (CPU and GPU parameters), game genres and themes, and game perspectives impact FPS. At the macro level, GDP per capita and the Gini index, representing average wealth and wealth distribution fairness respectively, indirectly affect players' average FPS performance in a given area~\cite{kashcha2022country, parshakov2018determinants}.

\begin{figure}[t]
\centering
\includegraphics[width=\textwidth]{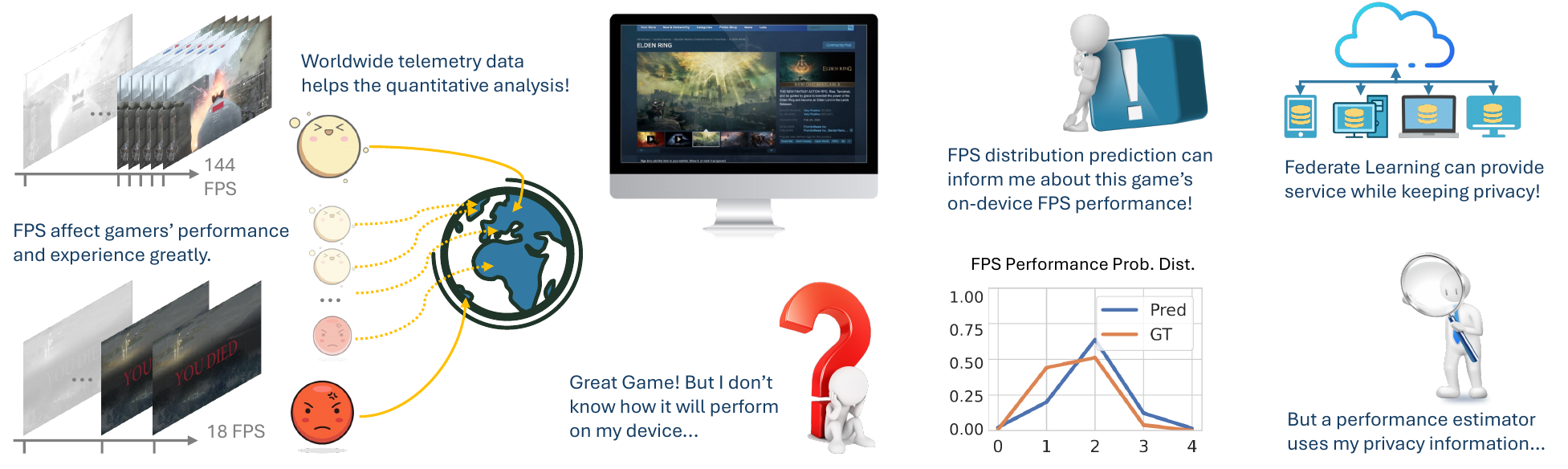}
\caption{The FPS of a game running on a device affect player's performance and mood. By analyzing telemetry data from all over the world, a federated learning-based FPS predictor is proposed for the convenience of players and game makers.}
\label{fig:cover}
\end{figure}

Given these determinants, training a predictive model requires sensitive information that players might be unwilling to share, such as hardware and software specifications and gaming session details. To address this concern, we propose a federated learning-based model that keeps private data on players' devices while still contributing to the global model.

Relying solely on a limited set of quantifiable features is insufficient for precise predictions, as many factors cannot be fully quantified. For example, dedicated gaming devices with minimal background programs and players who disable ray tracing may achieve higher FPS than others under identical conditions. Likewise, well-optimized games are more likely to reach higher FPS than less mature ones.

To better understand a player without collecting additional features, which could be intrusive, we propose assigning a unique learnable knowledge kernel (LKK) to each player. This knowledge kernel evolves as the player engages in more games, ultimately capturing their characteristics and modeling their behavior. Similarly, a game-specific LKK can also enhance game profiling.

While dedicated LKKs improve prediction accuracy, they require time to effectively profile a player or game. This leads to the "cold start problem," where kernels are not immediately applicable~\cite{lika2014facing, silva2023user}. To address this issue, we propose a unique training scheme and a dynamic switch mechanism for kernel application, which will be detailed in Section~\ref{methodology}.

\subsection*{Contributions}
% Our contributions through this research are as follows:\\
\begin{enumerate}
    \item We collected a gaming telemetry dataset encompassing 100,000 users from 224 countries and regions. This dataset contains records of 76.4 million game processes, including detailed software and hardware information and accurate real-time FPS performance data for 835 types of video game processes on various devices. It is the largest dataset of its kind globally.
    \item For games included in the telemetry dataset, we gathered 18 types of game characteristics, including genres, themes, and ratings, to model games from multiple perspectives.
    \item This study identifies potential factors affecting FPS and provides a comprehensive analysis of primary determinants on a global-scale dataset. It is the first study of its kind, offering valuable insights for data selection in future models.
    \item We developed a federated learning-based model to predict FPS distributions for games on new devices while addressing privacy concerns. Our model outperforms baseline methods, achieving a mean Wasserstein distance of 0.469 and a Cross Entropy of 1.3871 between predicted and ground truth FPS distributions.
    \item We introduced a novel training scheme that incorporates player- and game-specific learnable knowledge kernels to enhance prediction accuracy. This approach, which dynamically integrates these kernels, mitigates the cold start problem and results in a 7.57\% reduction in Wasserstein distance.
\end{enumerate}
 % and helping players make informed hardware upgrade decisions

% 1. We collected a gaming telemetry dataset encompassing 100,000 users from 224 countries and regions. This dataset contains records of 76.4 million game processes, including detailed software and hardware information and accurate real-time FPS performance data for 835 types of video game processes on various devices. It is the largest dataset of its kind globally.\\
% 2. We gathered 18 types of game characteristics, including genres, themes, and ratings, to model games from multiple perspectives.\\
% 3. This study identifies potential factors affecting FPS and provides a comprehensive analysis of primary determinants using a global-scale dataset. It is the first study of its kind, offering valuable insights for data selection in future models and helping players make informed hardware upgrade decisions.\\
% 4. We developed a federated learning-based model to predict FPS distributions for games on new devices while addressing privacy concerns. Our model outperforms baseline methods, achieving a mean Wasserstein distance of 0.469 and a Cross Entropy of 1.3871 between predicted and ground truth FPS distributions.\\
% 5. We introduced a novel training scheme that incorporates player- and game-specific learnable knowledge kernels to enhance prediction accuracy. This approach, which dynamically integrates these kernels, mitigates the cold start problem and results in a 7.57\% reduction in Wasserstein distance.

%% file: contents/2-related-work.tex
\section{Related Work} 

\subsection{Game Experience}

% The experience of playing video games is generally perceived as the personal connection between the player and the game, extending beyond the game's technical implementation. The experience is also regarded as a unique, personal engagement which involves both the process and the result of the interaction
% Player experience is constrained by the limitations of a games technical implementation. Processing power, graphic fidelity and frame rate, controller design, and user interface design all have a profound impact. 

The experience of playing video games is often seen as a personal connection between the player and the game, influnced by many factors~\cite{calvillo-gamezAssessingCoreElements2015, popperKnowledgeBodyMindProblem2013, mccarthy2004technology}. Researches into game experience have consistently identified several core factors that influence player satisfaction and engagement~\cite{claypoolEffectsFrameRate2006, claypoolPerspectivesFrameRates2009,liuEffectsFrameRate2023}. These factors include game mechanics~\cite{jagoda2018game,sanchez2012playability,moll2020players}, storyline, character design~\cite{soutter2016relationship,hefner2007identification,tompkins2022masculine}, soundtrack music~\cite{klimmt2019effects, lipscomb2004immersion}, control device~\cite{mcewan2012videogame}, and technical performance such as frame rate and resolution~\cite{claypoolEffectsFrameRate2006, claypoolPerspectivesFrameRates2009,liuEffectsFrameRate2023}. Although these papers provide some insightful points, their conclusion are either qualitative or based on a small group of players, lacking large data-driven worldwide analysis, and and statistics-based conclusion.

% \subsection{Frame Rate and Game Player’s Quality of Experience (QoE)}

\begin{figure}[h]
	\includegraphics[width=0.75\textwidth]{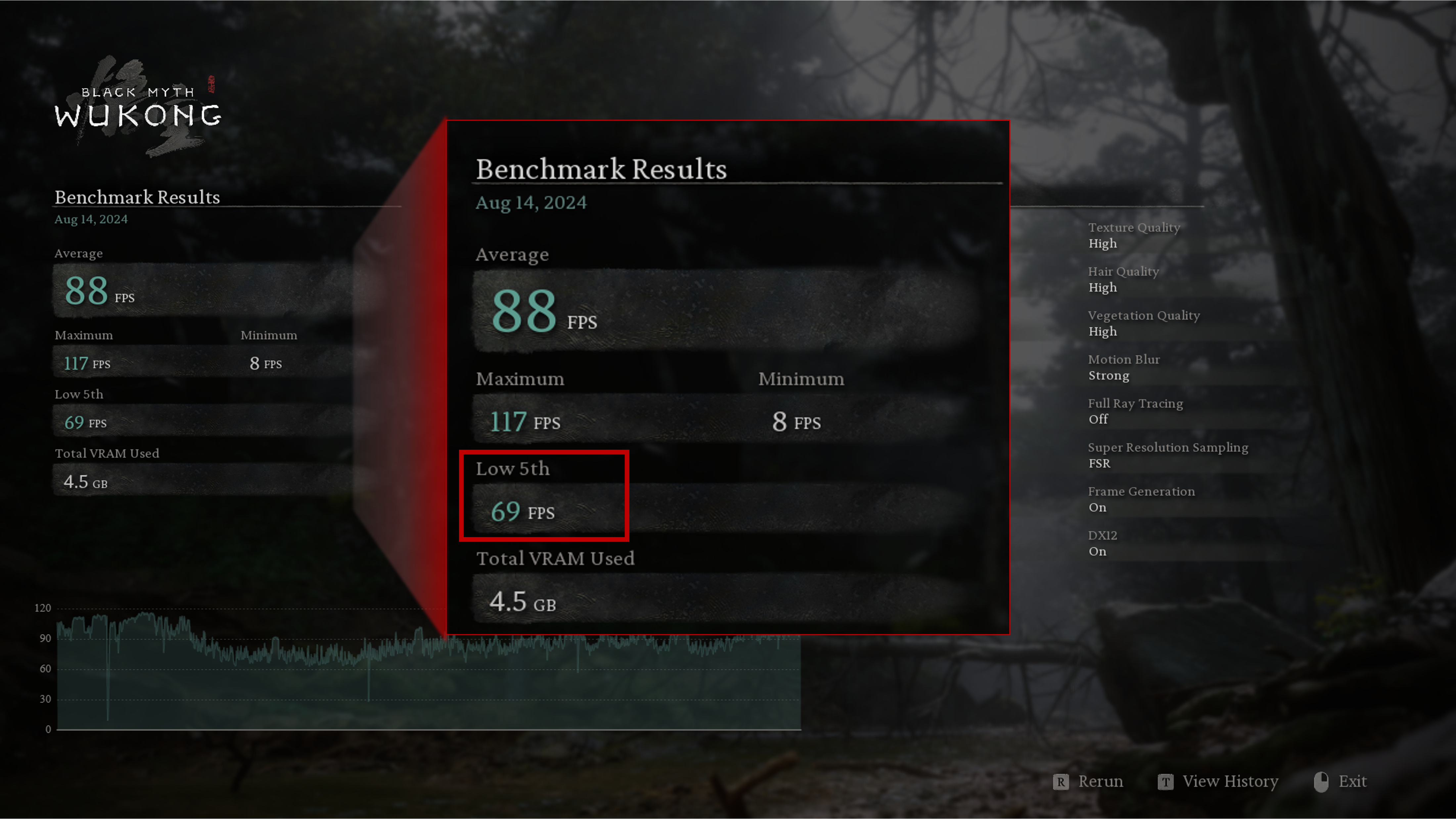}
 	\caption{The result of the game performance benchmark tool provided by Black Myth: Wukong, an AAA game released in August 2024. 95\% FPS floor is adopted as one of the main game performance metrics in this benchmark tool.} % in this dataset.
	\label{fig:wukong}
\end{figure}

Frame rate is crucial in shaping the Quality of Experience (QoE) for gamers~\cite{claypoolEffectsFrameRate2006, claypoolPerspectivesFrameRates2009}. Higher FPS generally contributes to smoother gameplay, reducing motion blur and input latency, which is critical for fast-paced games such as first-person shooters. Liu et al.~\cite{liuEffectsFrameRate2023} quantified the impact of frame rate variation on game player's quality of experience, noting that while average FPS is a general QoE predictor, it is not the best. The 95\% frame rate floor -- representing the bottom 5\% of frame rates experienced by the player during a gameplay -- effectively predicts QoE for both overall and individual games tested. This provides us with a strong conclusion that 95\% floor of FPS distribution is a good indicator of QoE, and our following analysis are done based on this conclusion. Recently, some top-tier games have started providing their own benchmark tools to calculate the 95\% floor FPS (also referred to as the ``Low 5th'') prior to players' purchase \cite{blackmythwukong}. They achieve this by asking players to pre-download a small demo-like sample game that encompasses most of the characteristic scenes from the main game. By automatically running this test tool for a period of time, it collects sufficient FPS sample data points and can provide the 95\% floor FPS metric. Although the performance test takes a relatively long time to complete, providing this type of benchmarking tool and informative metrics is largely welcomed by players. Fig. \ref{fig:wukong} illustrates an example of these tools.

Pioneering works have emphasized FPS as a key factor in the gaming experience, but fundamental influencing factors remain unexplored. Our quantitative analysis can help bridge this gap.

% We have addressed this gap by collecting a dataset that includes millions of players and billions of data points that reflect real-time FPS distributions during gameplay. This dataset was gathered while ensuring complete anonymity and without requiring demographic information such as gender. It encompasses various hardware and software details closely related to gaming, such as device type, GPU, CPU, screen size, game mode, fullscreen mode usage status, operating system, characteristics of over 800 frequently played games including game genre, ratings, popularity, and release date. Our subsequent quantitative analysis have yielded many interesting conclusions regarding FPS.

% \subsection{Game Performance Examination}

% Analyzing game performance involves assessing various system components, including CPU and GPU efficiencies, memory usage, and software optimization. XXXX outlined methodologies for measuring the impact of hardware and software configurations on game performance. Their research highlights the importance of optimized graphic settings and efficient coding practices to enhance FPS without sacrificing visual quality. Further, benchmarking tools and performance monitoring have become standard in evaluating the effectiveness of different hardware setups in running games at various graphical settings.

% \subsection{Federated Learning}
\subsection{Federated Learning with Distributed Data Sources}

Federated Learning (FL) is a transformative approach which enables model training across multiple decentralized devices~\cite{chenFSREALRealWorldCrossDevice2023,liReviewApplicationsFederated2020,mammenFederatedLearningOpportunities2021, pfeifferkilianFederatedLearningComputationally2023,10.1145/3381006}. Unlike traditional centralized training methods, FL allows data to remain on local devices: only client models' update or gradients are shared and aggregated. This approach not only reduces the risks and costs associated with data transmission but preserves client privacy. To enhance model performance, previous research has employed various techniques, including Federated Averaging (FedAvg) and its extensions~\cite{konecnyFederatedLearningStrategies2017,wangFederatedLearningMatched2020,karimireddySCAFFOLDStochasticControlled2020} to aggregate client model weights, and methods that aggregate gradients from clients~\cite{elbirFederated2020}.

% Federated Learning (FL) has emerged as a transformative approach in machine learning that enables the training of models across multiple decentralized devices while preserving data privacy~\cite{chenFSREALRealWorldCrossDevice2023,liReviewApplicationsFederated2020,mammenFederatedLearningOpportunities2021, pfeifferkilianFederatedLearningComputationally2023}. Unlike traditional centralized training methods, FL allows data to remain on local devices: only client models' update or gradients are shared and aggregated. This approach not only mitigates privacy concerns but also reduces the risks and costs associated with data transmission. The following studies have explored various aggregation techniques, including Federated Averaging (FedAvg) and its extensions~\cite{konecnyFederatedLearningStrategies2017,wangFederatedLearningMatched2020,karimireddySCAFFOLDStochasticControlled2020} which aggregate client model weights, and methods that aggregate client gradient instead~\cite{elbirFederated2020} to enhance model performance.

% Recent advancements in federated learning have addressed key challenges such as communication efficiency heterogeneity of data and devices, and security. 
% in diverse and non-IID (independent and identically distributed) settings~\cite{zhaoFederatedLearningNonIID2018}. 
Currently, FL has been broadly applied across domains including healthcare~\cite{riekeFutureDigitalHealth2020,stoffelFederatedLearningHealthcare2022,cFederatedLearningSmart2022}, finance~\cite{mammenFederatedLearningOpportunities2021,chatterjeeFederatedLearningEmpowered2024,awosikaTransparencyPrivacyRole2024}, and IoT~\cite{zhangFederatedFeatureSelection2023,issawaelBlockchainBasedFederatedLearning2023,venkatasubramanianIoTMalwareAnalysis2023}, demonstrating its potential to leverage large-scale distributed data sources while adhering to stringent privacy requirements. Inspired by previous works, we aim to leverage federated learning to
study player's video game experience with large-scale telemetry data.

% Furthermore, FL has found applications across various domains, including healthcare~\cite{riekeFutureDigitalHealth2020,stoffelFederatedLearningHealthcare2022,cFederatedLearningSmart2022}, finance~\cite{mammenFederatedLearningOpportunities2021,chatterjeeFederatedLearningEmpowered2024,awosikaTransparencyPrivacyRole2024}, and IoT~\cite{zhangFederatedFeatureSelection2023,issawaelBlockchainBasedFederatedLearning2023,venkatasubramanianIoTMalwareAnalysis2023}, demonstrating its potential to leverage large-scale distributed data sources while adhering to stringent privacy requirements. Inspired by previous works, we are the first to introduce federated learning to the study of video game experience with large-scale telemetry data.

% Inspired by these works, we are the first to introduce the federated learning mechanism to the video game experience study domain. With federated learning, our system does not need to upload any hardware or software information from players' devices yet can still predict the estimated FPS of a game on a specific device before the player purchases and installs it. This predictive model allows players to make informed purchasing decisions and helps game companies receive fewer negative comments about performance issues from players whose devices cannot run the game as expected.

% \subsection{How Existing Game Platforms Do}
\subsection{Game Performance and Configuration Guidance from Platforms}

% \begin{figure}[h]
%     \centering
%     \begin{subfigure}[t]{0.33\textwidth}
%         \centering
%         \includegraphics[height=1.6in, width=\textwidth]{images/RelatedWork/steam.eps}
%         \caption{Steam (minimum and recommended configuration).}
%         \label{fig:steam}
%     \end{subfigure}
%     \hfill
%     \begin{subfigure}[t]{0.33\textwidth}
%         \centering
%         \includegraphics[height=1.6in, width=\textwidth]{images/RelatedWork/Epic.png}
%         \caption{Epic GameStore (minimum and recommended configuration).}
%         \label{fig:epic}
%     \end{subfigure}
%     \hfill
%     \begin{subfigure}[t]{0.33\textwidth}
%         \centering
%         \includegraphics[height=1.6in, width=\textwidth]{images/RelatedWork/xbox.eps}
%         \caption{Xbox Games (performance prediction and configuration).}
%         \label{fig:xbox}
%     \end{subfigure}

%     \label{fig:steam-xbox}
% \end{figure}

\begin{figure}[h]
\includegraphics[width=\textwidth]{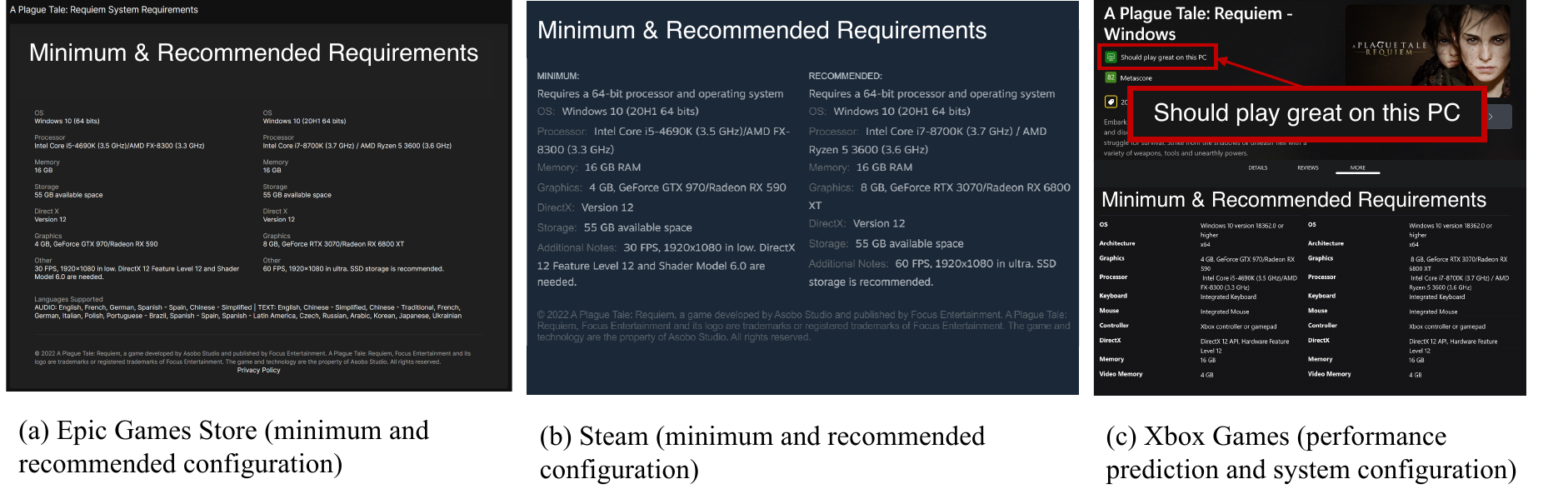}
\caption{Performance and configuration guide for the game \textit{``A Plague Tale: Requiem''} on (a) Epic GameStore, (b) Steam, and (c) Xbox Games. All game platforms provide minimum and recommended system configuration, while Xbox Games additionally provides data-driven binary running ability estimation.}
\label{fig:game-platforms}
\end{figure}

% \begin{figure}[h]
%     \centering
%     \begin{subfigure}[t]{0.33\textwidth}
%         \centering
%         \includegraphics[height=1.22in, keepaspectratio]{images/RelatedWork/steam.eps}
%         % \caption{Steam}
%         \caption{Steam (minimum and recommended configuration).}
%         \label{fig:steam}
%     \end{subfigure}
%     \hfill
%     \begin{subfigure}[t]{0.33\textwidth}
%         \centering
%         \includegraphics[height=1.22in, keepaspectratio]{images/RelatedWork/Epic.png}
%         % \caption{Xbox Games}
%         \caption{Epic GameStore (minimum and recommended configuration).}
%         \label{fig:xbox}
%     \end{subfigure}
%     \hfill
%     \begin{subfigure}[t]{0.33\textwidth}
%         \centering
%         \includegraphics[height=1.22in, keepaspectratio]{images/RelatedWork/xbox.eps}
%         % \caption{Xbox Games}
%         \caption{Xbox Games (performance prediction and configuration).}

%         \label{fig:xbox}
%     \end{subfigure}

%     % \caption{Game store screenshots of the same game ``A Plague Tale: Requiem'' on different game platforms.}
%     \caption{Performance and configuration guide for the game \textit{``A Plague Tale: Requiem''} on (a) Steam and (b) Xbox Games.}
%     \label{fig:steam-xbox}
% \end{figure}

Currently, mainstream gaming platforms like Steam~\cite{SteamStore}, Epic Games Store~\cite{EpicGamesStore2024}, and Xbox Games~\cite{XboxGames} on PC provide limited performance guidance, mainly listing minimum and recommended hardware configurations. While Epic Games Store (Fig.~\ref{fig:game-platforms}a) and Steam (Fig.~\ref{fig:game-platforms}b) only provide configuration requirements, Xbox Games (Fig.~\ref{fig:game-platforms}c) goes a step further by providing a brief note indicating whether the game ``should play great on this PC'' or not. But it doesn't offer more details like estimated average FPS or FPS distribution when playing this game.

However, these platforms still fall short in providing guidance from a frames-per-second (FPS) perspective, which could also consider a player's specific hardware and software characteristics, gaming history, and the performance of other users with similar setups. Our goal is to propose a more accurate and privacy-conscious solution to address this gap.

%% file: contents/3-dataset.tex
\section{Data Collection}

\begin{figure}[h]
	\includegraphics[width=0.9\textwidth]{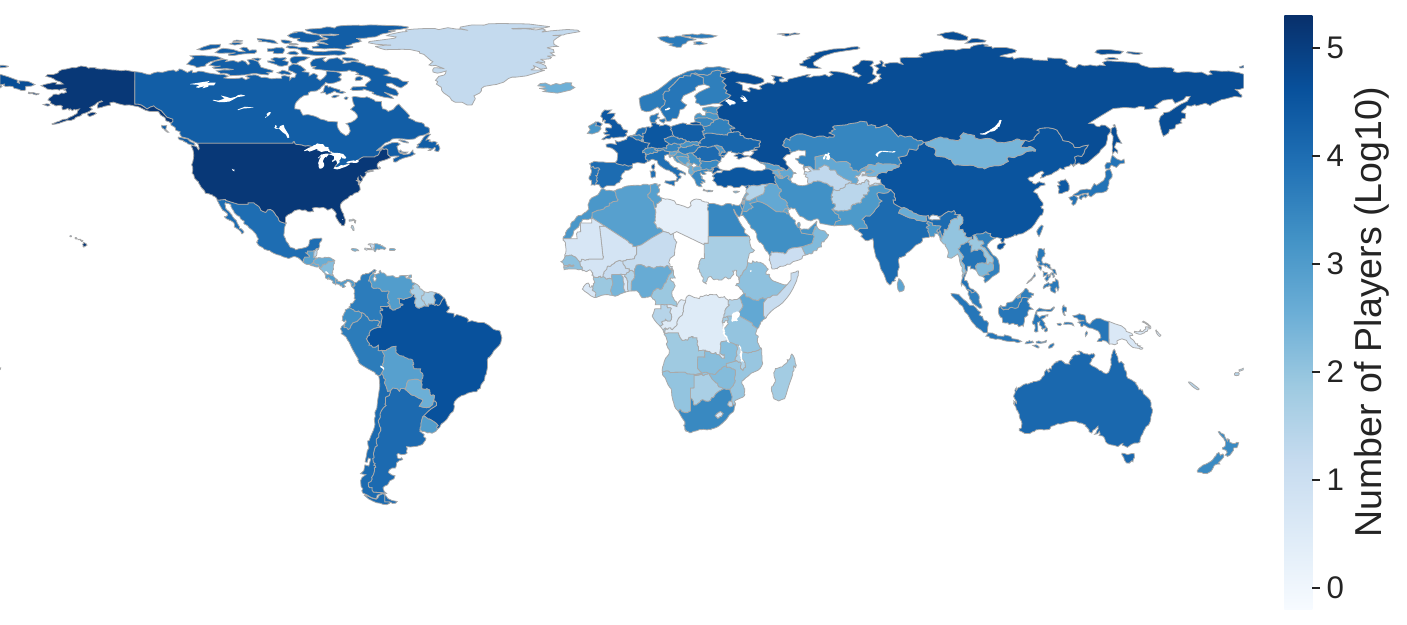}
	% \caption{The global player number in this telemetry dataset in log scale. Players from the United States, Brazil, Russia, China, and South Korea make up the largest share of players.} % in this dataset.
 	\caption{The global distribution of player number (log scale) in this telemetry dataset. The United States (18.87\%), Russia (6.98\%), Brazil (5.57\%), China (4.85\%), and Germany (4.27\%) make up the largest share of players.} % in this dataset.
	\label{fig:global_player_number}
\end{figure}

\sssec{Telemetry Data.} 
To better understand and predict what determines a player's FPS while playing a game, we designed a data collector. With user consent, this data collector runs in the background and primarily collects three types of data. 
\begin{enumerate}
    \item Player-specific data, such as the hardware device and operating system version used by the player, as well as the player's country or region (see Fig.~\ref{fig:global_player_number} for player distribution across different regions).
    \item Gameplay session-related data, including the game start time, whether it is in full screen mode, and whether Game Mode is enabled.
    \item FPS ground truth data. Once the game starts, our data collector samples and records the current frame rate every five seconds. Due to the varying lengths of game sessions, to reduce the volume of transmitted and recorded data and to avoid unnecessary precision, the data collector quantifies the FPS values into 42 bins. Each bin represents a 5Hz range, with four exceptions (<10Hz, 200-300Hz, 300-400Hz, >400Hz.) For each game session, it records the number of FPS samples in each bin and the average FPS for the session. Data collector on each device is assigned a unique identifier to tell apart players.
\end{enumerate}
  
Since we aim to collect only game-related processes' data and determine whether a Windows process is a game program, we filter processes depending on if it invokes DirectX~\cite{DirectX}. Microsoft DirectX is a collection of application programming interfaces (APIs) for handling tasks related to multimedia, especially game programming and video, on Microsoft platforms. When a program calls DirectX while in the foreground, it triggers our data collector to generate a new record.

To further collect game-related data and exclude non-game processes, we mapped executable names to games by web scraping SteamDB~\cite{SteamDB}, a third-party site listing executable names under each game's configuration page~\cite{GrandTheftAuto}. For executables not mapped from the scraped data, we manually annotated their game names through web searches and individual checks (e.g., mapping BlackOps.exe'' to \textit{Call of Duty: Black Ops''}). Additionally, executable names that are too ambiguous to identify specific games (e.g., game.exe'', launcher.exe'') were discarded along with non-game processes.

We found that some users had not played any games, with records triggered by other programs invoking DirectX. We removed these sessions from our dataset. We also excluded game sessions shorter than 5 minutes, as initial welcome pages and game loading times distort the actual FPS distribution. Players who initiated game processes fewer than 18 times were also excluded due to insufficient FPS samples for meaningful probability analysis (Section~\ref{sec:fps-pred}). 

% For more details about the exact feature types and explanations, please refer to Table 1 in the supplementary material.

\sssec{Game Characteristics Data.} 
The telemetry data does not reveal game characteristics-related features. However, these features could be crucial confounding factor when it comes to FPS prediction~\cite{claypoolEffectsFrameRate2006}. To address this gap, we incorporated broader game context and attributes from external sources. Following the executable-game name mapping process mentioned above, we collected game characteristics data from IGDB~\cite{IGDBComDiscover}, a platform offering extensive information on games across various platforms. We gathered details such as genres, types, perspectives, age ratings, and game rating conditions. A complete list of the selected features and their explanations is provided in the supplementary materials.

% The complete list of all the feature names and explanations are presented in the Table 2 in the supplementary material.
% details or characteristics including game genre, type, age rating, popularity, and rating conditions

% for the complete selection of features, see Table~\ref{tab:gamefeat}

For our analysis, games that are not well represented in the telemetry dataset (i.e., has less than 10 records) are excluded from the game characteristics dataset. For some niche games which were either not listed in IGDB or had significant amount of missing features, we filtered them out with careful manual check. Ultimately, we selected 835 games in our game characteristics data for subsequent analysis and model training. Among these, 153 games had at least 25 players in our telemetry dataset.

\sssec{Country-Level Data.}
% \jinhe{may reframe this part, refer to the parts using this data in the data insight section}
%The aforementioned two datasets are instrumental in studying the relationships between FPS and hardware, software, and game characteristics. While these factors directly impact FPS, 
We are also interested in the macro-level factors influencing FPS, the proxy variable of interest representing game experience. After filtering, we found a strong correlation between FPS and the overall level of economic development and income equality of a country or region.

To quantitatively study these relationships, we obtained the 2018 GDP (Gross Domestic Product) per capita data for various countries from the IMF website~\cite{IMFData} and downloaded the Gini index of the same year from the World Bank database~\cite{WorldBank}. For countries or regions with missing data for 2018, we used the closest available data. If no data could be found within a decade, the country was excluded from the macro-level analysis.

%% file: contents/4-data-insights.tex
\section{Data Insights}
% As previously discussed, t
% The 95\% FPS floor is a key metric for evaluating a player's experience in a video game~\cite{liuEffectsFrameRate2023}. In this section, we will introduce and thoroughly explain the factors associated with this metric.
The 95\% FPS floor is a key metric to evaluate players' quality of game experience~\cite{liuEffectsFrameRate2023}. In this section, we provide an in-depth analysis to uncover and explain the factors associated with this metric.

% As discussed previously, the 95\% FPS floor is an key metric for assessing a player's experience while engaging with a video game~\cite{liuEffectsFrameRate2023}. In this section, we will introduce these influencing factors and explain them in depth. 
% In this section, we call ``95\% FPS floor'' as ``experience indicator'' for explanation convenience.

\subsection{Micro-Level Influence Factors}

% %%%%%%%% FULL DEVICES ANOVA TABLE %%%%%%%%%%%
% \begin{table*}[]
% \caption{One-way ANOVA test results on some categorical device-related features. Samples with missing values on specific features are excluded from the corresponding test. ``SSW'' and ``SSB'' stand for ``Sum of Squares Within/Between Groups''.}
% \begin{tabular}{lllrrrrr}
% \toprule
%          Feature &       SSB &       SSW &  DF Between &  DF Within &   F Value &  p Value &  $\eta^2$ \\
% \midrule
%      Device Type & 1.862e+07 & 1.426e+08 &           5 &     153249 &  4003.460 &      0.0 &     0.116 \\
% Operating System & 1.111e+06 & 1.602e+08 &           5 &     153370 &   212.757 &      0.0 &     0.007 \\
%      % Screen Size & 2.084e+07 & 9.432e+07 &          13 &     119984 &  2038.940 &      0.0 &     0.181 \\
%       CPU Family & 1.681e+07 & 1.333e+08 &           8 &     146876 &  2314.744 &      0.0 &     0.112 \\
%  % CPU Core Number & 4.298e+07 & 1.072e+08 &          13 &     146867 &  4531.101 &      0.0 &     0.286 \\
%        GPU Class & 4.087e+07 & 1.181e+08 &           3 &     150979 & 17422.867 &      0.0 &     0.257 \\
%     GPU Category & 2.897e+07 & 1.299e+08 &           3 &     151007 & 11230.100 &      0.0 &     0.182 \\
%          Country & 7.933e+06 & 1.494e+08 &         198 &     149293 &    40.049 &      0.0 &     0.050 \\
% \bottomrule
% \end{tabular}
% \label{tab:anova-device}
% \end{table*}
% %%%%%%%% FULL DEVICES ANOVA TABLE %%%%%%%%%%%

% \subsubsection{Hardware}

\sssec{Player-side features.} We examined the factors influencing the 95\% FPS floor, focusing on representative player characteristics like CPU and GPU specifications, operating system, country and region, and device type.

% (1) \textit{Device type.} 

\subsubsection{Device type.} Different devices have distinct design philosophies and trade-offs. For example, tablets prioritize portability but often sacrifice performance and cooling, while desktops, though bulkier, can house more powerful components and benefit from a stable power supply due to their larger size. We categorized devices into six types: desktop, notebook, 2-in-1 devices, Intel NUC/STK (Next Unit of Computing/Compute Stick), server/workstation, and tablet. To quantitatively analyze how device types affect the 95\% FPS floor, we conducted an one-way ANOVA test~\cite{Quirk2012oneway} comparing the effects of different device types. The results, presented in Table~\ref{tab:anova-device}, revealed a statistically significant difference in the 95\% FPS floor between at least two groups ($F(6, 153387) = 4003.460, p = 0$), with a medium effect size ($\eta^2 = 0.116$). Furthermore, the Tukey’s HSD test~\cite{TukeyMethod} for multiple comparisons found significant differences between groups such as desktop and 2-in-1 devices ($p=0, 95\% C.I. = [31.847, 34.638]$), and server/workstation and tablet ($p=0.024, 95\% C.I. = [-74.527, -3.029]$). However, there was no statistically significant difference in the mean 95\% FPS floor between notebook and server/workstation ($p=0.558$) or 2-in-1 devices and tablets ($p=0.585$).

%%%%%%%% PART DEVICES ANOVA TABLE %%%%%%%%%%%
\begin{table}[ht]
\caption{One-way ANOVA test results on categorical device features. Samples with missing values on specific features are excluded from the corresponding test. ``DFW/DFB'' stand for ``Degrees of Freedom Within/Between Groups''.}
\begin{tabular}{lrrrrr}
\toprule
         Feature &     DFB &  DFW &   F &  p &  $\eta^2$ \\
\midrule
     Device Type &     5 &     153249 &  4003.460 &      0.000 &     0.116 \\
Operating System &     5 &     153370 &   212.757 &      0.000 &     0.007 \\
  % Screen Size  &    13 &     119984 &  2038.940 &      0.000 &     0.181 \\
      CPU Family &     8 &     146876 &  2314.744 &      0.000 &     0.112 \\
% CPU Core Number&    13 &     146867 &  4531.101 &      0.000 &     0.286 \\
        GPU Level &   4  &     150978 & 13221.208 &      0.000 &     0.259 \\
       % GPU Level &     3 &     150979 & 17422.867 &      0.000 &     0.257 \\
    GPU Category &     3 &     151007 & 11230.100 &      0.000 &     0.182 \\
         Country &   198 &     149293 &    40.049 &      0.000 &     0.050 \\
\bottomrule
\end{tabular} 
\label{tab:anova-device}
\end{table}

% \begin{table}[]
% \centering
% \caption{Summary of Tukey's HSD test for pairwise comparisons of device type.}
% \begin{tabular}{llrrrr}
% \toprule
%        group1 &        group2 &  meandiff &  p-adj &   lower &   upper \\
% \midrule
%        2 in 1 &       Desktop &    33.243 &  0.000 &  31.847 &  34.638 \\
%        2 in 1 & NUC/STK &    16.908 &  0.000 &  13.073 &  20.742 \\
%        2 in 1 &      Notebook &    11.858 &  0.000 &  10.471 &  13.246 \\
%        2 in 1 &     Server/WS &    20.077 &  0.001 &   5.913 &  34.242 \\
%        2 in 1 &        Tablet &   -18.701 &  0.585 & -51.580 &  14.178 \\
%       Desktop & NUC/STK &   -16.335 &  0.000 & -19.938 & -12.732 \\
%       Desktop &      Notebook &   -21.384 &  0.000 & -21.838 & -20.930 \\
%       Desktop &     Server/WS &   -13.165 &  0.083 & -27.269 &   0.938 \\
%       Desktop &        Tablet &   -51.943 &  0.000 & -84.796 & -19.090 \\
% NUC/STK &      Notebook &    -5.049 &  0.001 &  -8.650 &  -1.449 \\
% NUC/STK &     Server/WS &     3.170 &  0.990 & -11.379 &  17.718 \\
% NUC/STK &        Tablet &   -35.608 &  0.026 & -68.655 &  -2.562 \\
%      Notebook &     Server/WS &     8.219 &  0.558 &  -5.884 &  22.322 \\
%      Notebook &        Tablet &   -30.559 &  0.085 & -63.412 &   2.293 \\
%     Server/WS &        Tablet &   -38.778 &  0.024 & -74.527 &  -3.029 \\
% \bottomrule
% \end{tabular}
% \label{tab:device-hsd}
% \end{table}

% The one-way ANOVA revealed that there was a statistically significant difference in 95\% FPS floor between at least two groups ($F(4, 150978) = 13221.208, p = 0$) ($F(3, 151007)=11230.100, p=0$),

\subsubsection{GPU Performance.} Delving into specific hardware configurations, we find that GPU performance is closely related to FPS, as shown in Fig.~\ref{fig:gpu_level}. Given that GPU performance is influenced by many parameters, we grouped GPUs with publicly available specifications into five categories: Office-Class, Low-End, Low-Mid-Range, Mid-Range, and High-End. The classification details are available on the GPU ranking website~\cite{hinumMobileGraphicsCards}. Records with missing GPU model values or unknown GPU types were excluded. A one-way ANOVA revealed a statistically significant difference in the 95\% FPS floor between at least two groups, as shown in Table~\ref{tab:anova-device} and Fig.~\ref{fig:gpu_level}. FPS distributions also significantly differed between players using discrete Nvidia, AMD, and Intel graphics cards or integrated graphics. These results are further detailed in Table~\ref{tab:anova-device}, with the corresponding box plot shown in Fig.~\ref{fig:gpu_category}. This variation is likely due to differences in game optimization across GPU brands and exclusive technologies like Nvidia's DLSS (Deep Learning Super Sampling), which can achieve higher frame rates with similar computational power~\cite{watsonDeepLearningTechniques2020}. 

\subsubsection{CPU Performance.} 
\begin{figure}[ht]
    \centering
    \begin{subfigure}[b]{0.6\textwidth}
        \centering
        \includegraphics[width=\textwidth]{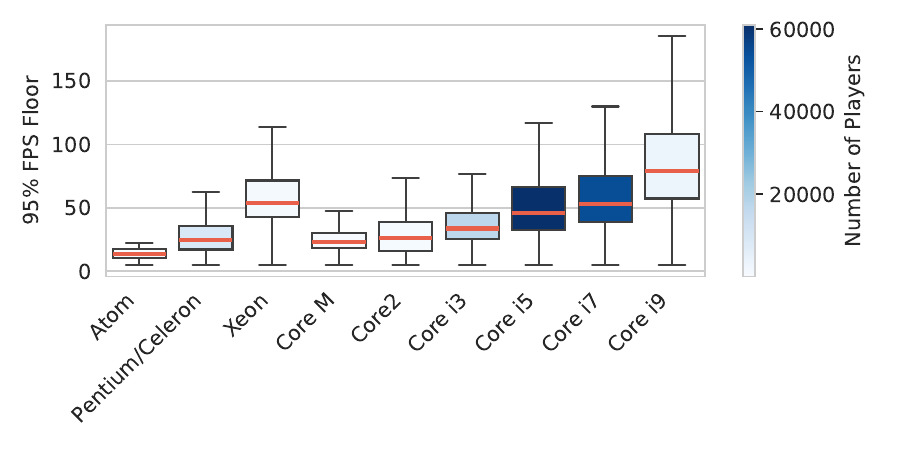}
        \caption{CPU family.}
        \label{fig:cpu_family}
    \end{subfigure}
    % \hspace{0.01\textwidth}
    \begin{subfigure}[b]{0.3\textwidth}
        \centering
        \includegraphics[width=\textwidth]{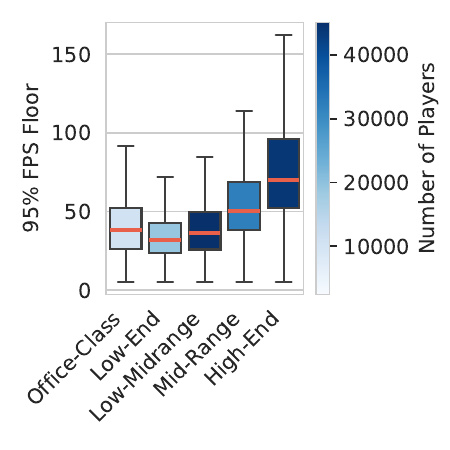}
        \caption{GPU Level.}
        \label{fig:gpu_level}
    \end{subfigure}
    
    \vspace{0.05\textwidth} % Vertical space between the rows

    \begin{subfigure}[b]{0.6\textwidth}
        \centering
        \includegraphics[width=\textwidth]{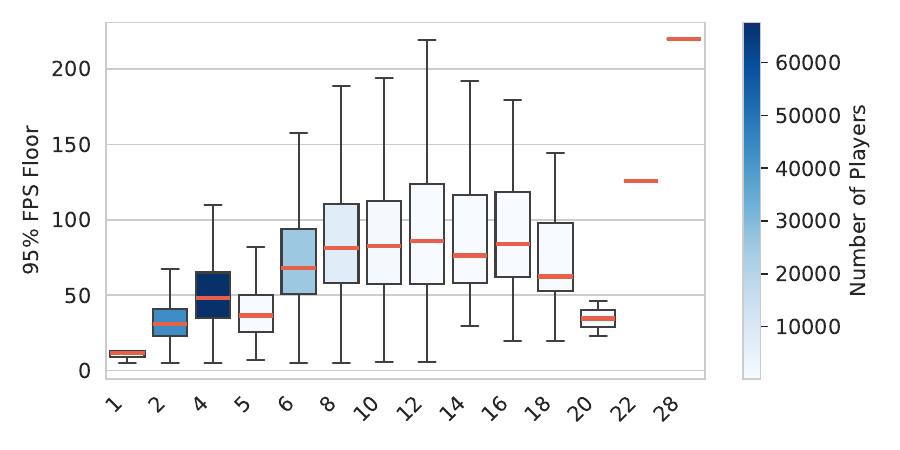}
        \caption{CPU Core Number.}
        \label{fig:cpu_ncores}
    \end{subfigure}
    \hspace{0.05\textwidth}
    \begin{subfigure}[b]{0.3\textwidth}
        \centering
        \includegraphics[width=\textwidth]{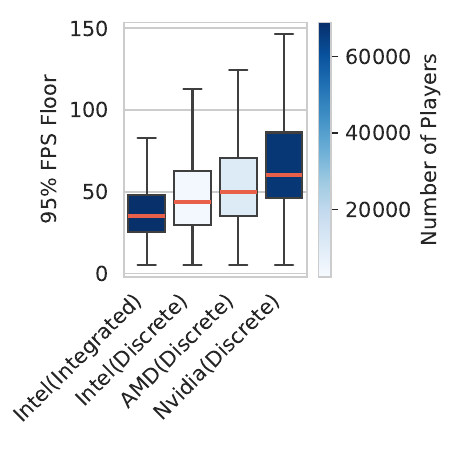}
        \caption{GPU Category.}%Manufacturer.}
        \label{fig:gpu_category}
    \end{subfigure}
    
    \caption{The distribution of 95\% FPS floor across CPU and GPU core parameters. Higher-spec GPU and CPU directly lead to a higher 95\% FPS floor.}
    \label{fig:gpucpu}
\end{figure}

Similarly, CPU performance positively affects FPS, with improvements in CPU generation (Fig.~\ref{fig:cpu_family}) and core count (Fig.~\ref{fig:cpu_ncores}) resulting in a higher 95\% FPS floor. One-way ANOVA tests conducted for CPU families show a significant effect, as detailed in Table~\ref{tab:anova-device}. A multiple linear regression was used to test if CPU process node (unit: nm), CPU core count, RAM size, and screen size significantly predicted the 95\% FPS floor. The fitted regression model was:
\begin{equation}
    FPS_{95} = 6.878 \times n + 0.28 \times r  - 0.5165 \times p + 1.8697 \times s - 2.0675
\end{equation}
, where $n$, $r$, $p$, and $s$ represent the CPU core count, RAM size (in GB), CPU process node (in nm), and screen size (in inches), respectively. The model explained approximately 33.3\% of the variance in the dependent variable, as indicated by an $R^2$ value of 0.333. More details are provided in Table~\ref{tab:lr-device}. While a larger screen size doesn't intuitively lead to better FPS, higher investment in peripherals like monitors often reflects a player's greater demand for overall gaming experience. Such players are likely to invest more in core device components, which may explain this correlation.

%%%%%%%% PART DEVICES ANOVA TABLE %%%%%%%%%%%

\begin{table}[ht]
\caption{Summary of multiple linear regression analysis for device-related features. ``Lower/Upper'' stand for ``Lower/Upper Confidence Interval Bound''}
\begin{tabular}{lrrrrr}
\toprule
        Feature &  Coef. &  p &  $R^2$ &  Lower &  Upper \\
\midrule
  RAM Size (GB) &       0.2843 &      0.000 &      0.333 &            0.262 &            0.307 \\
    CPU Process Node &      -0.5165 &      0.000 &      0.333 &           -0.540 &           -0.493 \\
CPU Core Number &       6.8780 &      0.000 &      0.333 &            6.770 &            6.986 \\
    Screen Size (inch) &       1.8697 &      0.000 &      0.333 &            1.825 &            1.914 \\
\bottomrule
\end{tabular}
\label{tab:lr-device}
\end{table}

% which in turn demands more from core hardware components like the GPU and CPU, typically resulting in higher FPS.
% players willing to invest more in peripherals like screen is more likely to invest more in other core components as well, which can explain this correlation. 

\subsubsection{Operating System.} In addition to hardware specifications, we also examined the impact of the operating system. The corresponding one-way ANOVA results ($F(5, 153370) = 212.757, p = 0$) showed a statistically significant difference between groups, though the effect size was very small ($\eta^2 = 0.007$), suggesting that these differences, while statistically significant, have limited practical importance.

\subsubsection{Country and Region.} Finally, we explored the effect of the player's country or region. As shown in Table~\ref{tab:anova-device}, there was a significant difference between country groups, with a small effect size ($\eta^2 = 0.05$). Country-level impact factors are discussed in detail in Section~\ref{sec:macro}.

\begin{table}[ht]
\centering
\caption{One-way ANOVA test results on categorical game features. The degree of freedom is not static since one game can have multiple tags. This table shows that game characteristics are not as influential to 95\% FPS floor as device features.}
\begin{tabular}{lrrrrr}
\toprule
            Features &  DFB &  DFW &  F &  p &  $\eta^2$ \\
\midrule
             Themes &     21 &  1822 &    2.120 &    0.002 &     0.024 \\
             Genres &     22 &  2232 &    2.346 &    0.000 &     0.023 \\
Player Perspectives &      6 &        903 &    4.412 &    0.000 &     0.028 \\
         Game Modes &      5 &       1695 &    1.064 &    0.379 &     0.003 \\
   DLCs Availablity &      1 &        833 &    0.213 &    0.644 &     0.000 \\
        Age Ratings &      1 &        668 &    0.753 &    0.386 &     0.001 \\
\bottomrule
\end{tabular}
\label{tab:anova-game}
\end{table}
%%%%%%%% FULL DEVICES ANOVA TABLE %%%%%%%%%%%

\begin{table}
\caption{Summary of multiple linear regression analysis for game characteristic features.  $^\dagger$``Lang. Sup.'' is the abbreviation of ``Languages Supported.''}
\begin{tabular}{lrrrrr}
\toprule
                  Features &  Coef. &  p &  $R^2$ &  Lower &  Upper \\
\midrule
Platforms Sup.$^\dagger$ &      17.099 &   0.025 &      0.011 &            2.138 &           32.061 \\
    Release Year &      -6.576 &   0.466 &      0.011 &          -24.271 &           11.119 \\
    Release Month &      -3.924 &   0.181 &      0.011 &           -9.675 &            1.826 \\
Lang. Sup.$^\dagger$ &      -0.754 &   0.884 &      0.011 &          -10.903 &            9.394 \\
Rating &      10.212 &   0.256 &      0.011 &           -7.428 &           27.852 \\
Rating Count &      -5.132 &   0.568 &      0.011 &          -22.765 &           12.502 \\
\bottomrule
\end{tabular}
\label{tab:lr-game}
\end{table}

\sssec{Game Characteristics.} Beyond hardware settings, the impact of game characteristics cannot be overlooked. We also analyzed categorical and numerical features separately. 

\subsubsection{Categorical game features.} One-way ANOVA results revealed a significant difference in the 95\% FPS floor across different game themes ($F(21, 1822) = 2.12, p = 0.002, \eta^2 = 0.024$), though the effect size was small. Similarly, game genres and player perspectives (first/third person, bird's-eye view, etc.) also showed a small effect on the 95\% FPS floor, as indicated by the ANOVA results in Table~\ref{tab:anova-game}. The table also shows that other game characteristics, such as game modes (single/multiplayer, battle royale, etc.), DLC (Downloadable Content) availability, and age ratings (Either 18+ or not), do not exhibit significant differences between groups. It's worth noting that one game can have multiple tags for a categorical feature, so the degree of freedom change accordingly.

\subsubsection{Numerical game features.} Additionally, a multiple linear regression was performed to assess whether numerical game characteristic features significantly predicted the 95\% FPS floor. The results, shown in Table~\ref{tab:lr-game}, indicate that the estimated coefficient for the number of supported game platforms was 17.099, with a significant p-value of 0.025, suggesting that this feature has a meaningful effect on the 95\% FPS floor. However, p-values for game release years and months, game rating, rating count, and supported language count were all greater than 0.05, indicating no statistically significant relationship with the 95\% FPS floor. This suggests that these game characteristics do not have a strong or consistent effect on the 95\% FPS floor within the context of this dataset.

Compared to player-side features like device parameters, game characteristics have a significantly smaller effect on the 95\% FPS floor, as observed from both the F-statistic and $\eta^2$ results in Table~\ref{tab:anova-device} and Table~\ref{tab:anova-game}. This indicates that hardware specifications have a greater impact on game experience than game characteristics.

\subsection{Macro-Level Influence Factors}\label{sec:macro}
The previous sections discussed micro-level factors affecting FPS performance. But on a macro level, what broader factors influence the gaming experience?

\subsubsection{GDP per Capita.} Our analysis reveals a positive correlation between a country's average gaming experience and its GDP per capita, following a logarithmic relationship ($p = 0, R^2 = 0.562$). GDP per capita reflects economic output and living standards. Countries with high GDP per capita, such as Iceland and Switzerland, show the highest FPS performance, while those with low GDP per capita, such as Rwanda and Togo, exhibit the lowest. The scatter plot and fitted line are shown in Fig.~\ref{fig:95p-gdp}, with some highlighted countries.

%($p=6.64e^{-27}$, $R^2=0.562$)

\subsubsection{Gini index.} 
\begin{figure}[ht]
    \centering
    \begin{subfigure}[b]{0.6\textwidth}
    \includegraphics[width=\textwidth]{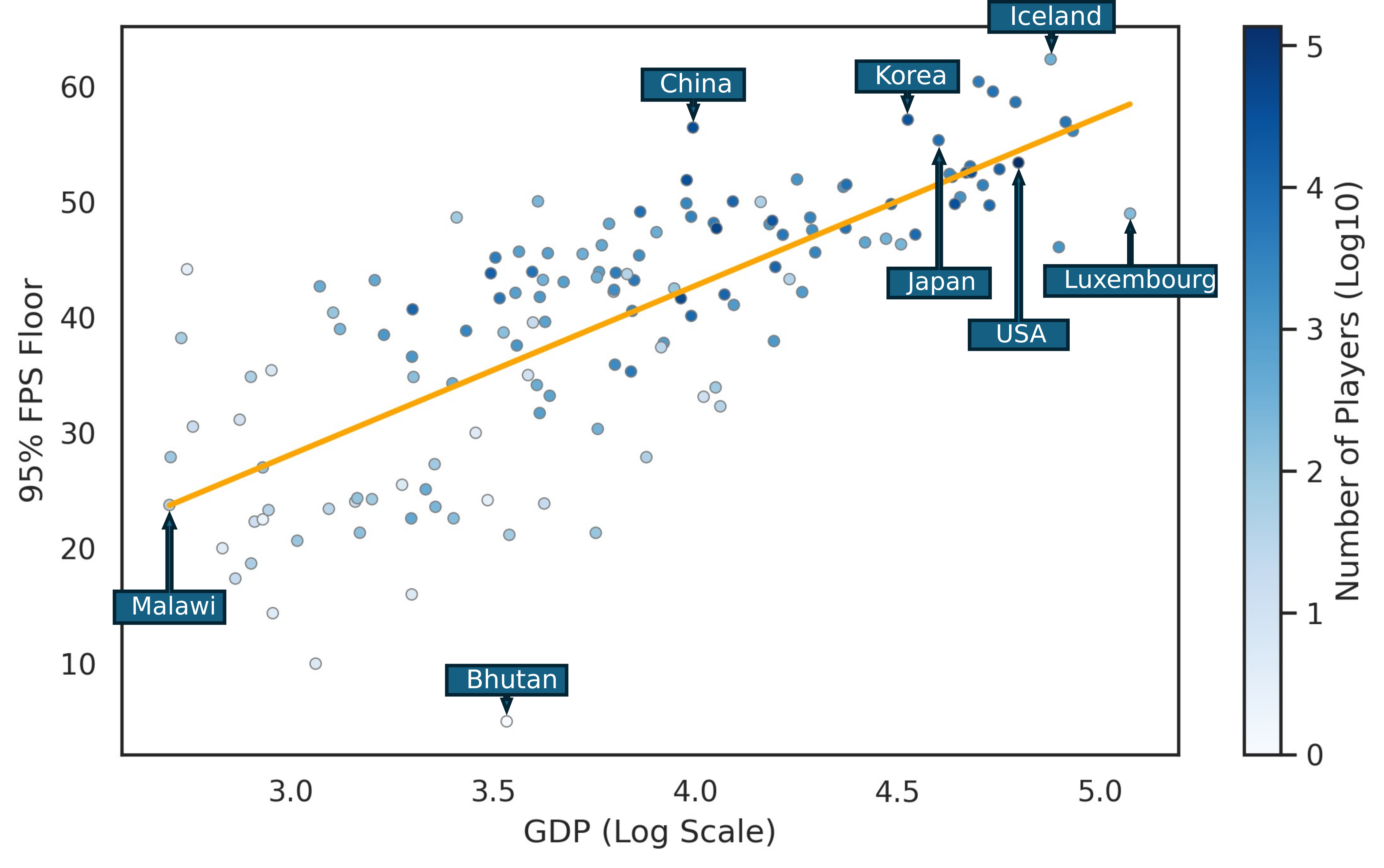}
    \caption{Scatter plot of 95\% FPS floor vs. GDP per capita showing logarithmic positive correlation.}
    \label{fig:95p-gdp}
    \end{subfigure}
    \hfill
    \begin{subfigure}[b]{0.6\textwidth}
        \includegraphics[width=\textwidth]{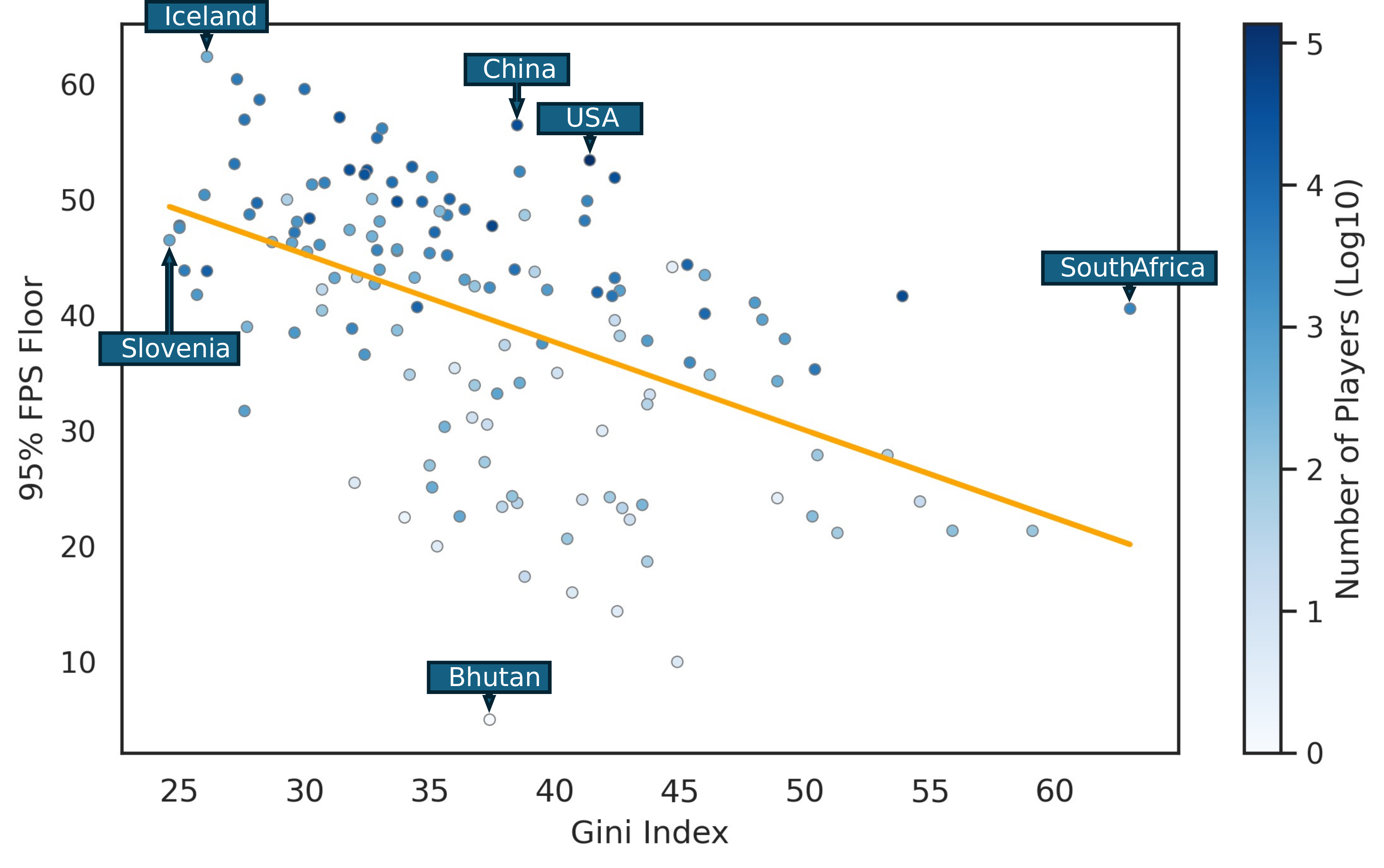}
        \caption{Scatter plot of 95\% FPS floor vs. Gini Index showing negative correlation.}
        \label{fig:95p-gini}
    \end{subfigure}
    \caption{The 95\% FPS floor's correlation with countries' GDP per capita (Log Scale) and Gini index.}
    \label{fig:95p-gdp-gini}
\end{figure}

Conversely, the Gini index, which measures income inequality, negatively correlates with gaming experience ($p = 0, R^2 = 0.246$). Higher Gini indices, indicating greater inequality, are associated with lower gaming experiences. Countries with low Gini indices, such as Iceland and Denmark, have the best player experiences, while those with high Gini indices, like Namibia and Colombia, exhibit poorer performances. Corresponding figure is shown in Fig. \ref{fig:95p-gini}.
%($p=3.38e^{-10}$, $R^2=0.246$)

By fitting a line to our data, we derived a simple formula that accurately describes the relationship between a country's average 95\% FPS floor, GDP per capita, and the Gini Index:
\begin{equation}
    FPS_{95} = 12.84 \times log_{10}\alpha + -0.42\beta + 6.84
\end{equation}
, where $\alpha$ and $\beta$ represents GDP per capita (in USD)  and the Gini Index of the country or region, respectively. When these two factors are considered together, the model yields an R-squared value of 0.625, indicating that approximately 62.5\% of the variability in the 95\% FPS floor is explained by variations in GDP and Gini Index. The statistical significance of these predictors is confirmed by the resulting p-value ($p = 0$).
% , demonstrating their significant impact on the 95\% FPS floor. ($8.72e^{-31}$)

In summary, on a micro level, better hardware such as GPUs, CPUs, and RAM, as well as performance-focused devices, contribute to an overall better gaming experience. Game characteristics, including themes, genres, and player perspectives, also play a role. On a macro level, the average wealth of a population and the fairness of income distribution within a country are significant factors influencing the gaming experience.

% In summary, on a micro level, better hardwares like GPUs, CPUs, and RAMs, high-performance type of devices contribute to an overall better gaming experience. Game characteristics like themes, genres, and player perspectives also have an impact. On a macro level, the average wealth of a population and the fairness of income distribution within a country are significant factors influencing the gaming experience.

% approximately 62.5\% of the variability in 95\% FPS floor is explained by the variations in GDP per capita and the Gini index, .

% In summary, on a micro-level, better GPUs and CPUs, more suitable manufacturers, and performance-centric devices all lead to an overall better gaming experience. Types and genres of games also cast an influence. On a macro level, the average wealth of people and the fairness of the distribution within a country are influential factors.

% \subsection{The Impact of FPS}
% Previously, we discussed the factors that determine the user experience when the 95\% FPS floor is used as a metric. However, FPS is not merely an endpoint of exploration; it also profoundly influences various aspects related to gaming.

% First, it affects the lifecycle of a game.

% Second, it impacts the average duration of each gaming session.

% Furthermore, FPS influences the frequency with which players engage in gaming.

% From the above, it is clear that FPS is an excellent metric for measuring gaming experience, which allows us to somewhat characterize player behavior and thus guide more personalized game design.

%% file: contents/5-predictor.tex
\section{Player-Game Pair FPS Distribution Predictor}\label{sec:fps-pred}

Building on top of the statistical relations we found in the previous section, there is a clear correlation between FPS distribution and player and game features. In this section, we aim to develop a predictive model for game experience with telemetry data. 

% This predictor is beneficial to players, game developers, and gaming platforms.

% \sssec{For players: }
% FPS affects players' gaming performance and quality of experience. If the game cannot achieve the minimal tolerable FPS on a player's device, he or she is highly likely to hold negative opinion towards the game or give it up. Having an accurate FPS distribution estimation is essential to help players make wise and informed decision.

% \sssec{For Game Developers: }
% On platforms like Steam, poor game optimization and low FPS often lead to player dissatisfaction and commercial setbacks, as seen with titles like ``Cyberpunk 2077'' \cite{cyberpunk2077} and ``Nobunaga’s Ambition'' \cite{Nobunaga}. While enhancing game quality is ideal, letting potential customer's know about the estimated FPS performance can help with expectation management and mitigate negative reviews and revenue loss.

% \sssec{For Gaming Platforms: }
% Accurate FPS performance predictions can enhance platform service by better aligning game offerings with user hardware and habits. This proactive approach not only boosts customer satisfaction but also supports game developers by refining performance insights and consulting services.

% In summary, integrating FPS performance predictors benefits players, developers, and platforms, aiming to improve the gaming experience and industry standards.

\subsection{Data Pre-processing}

The collected telemetry and game-centric data includes various data types, such as floating-point, integer, categorical, date-time, boolean, and string. Data preprocessing involved cleaning and transforming raw IGDB data into a structured format, addressing missing values by filling numerical ones with the median and normalizing them, while categorical features were one-hot embedded and missing values were set to 0.5. This preparation ensured a consistent and analyzable dataset for comprehensive analysis and modeling.

Additionally, the basic unit of data used in the subsequent deep learning network is the ``player-game pair,'' which represents the FPS performance probability distribution for each player when playing each game. For each game played by a player, a record is created, which includes relevant player information and game characteristics. After obtaining the raw telemetry data, we performed a data merge based on player-game pairs. For numerical features, we calculated the average value; for categorical and boolean features, we determined the mode; and for each FPS bin, we summed the values directly.

FPS is an instantaneous value, and it is impossible for a game running on a specific device to maintain a fixed FPS forever. Due to varying game scenes that demand different computational resources, an FPS distribution more accurately profiles a game's actual performance on a particular device.
To establish ground truth, we grouped the collected FPS values into five classes using thresholds of 25Hz, 45Hz, 60Hz, and 145Hz. These thresholds are used to define gameplay smoothness, aligning with common standards to cover diverse game scenarios and increase robustness. However, it is important to note that while these aggregated FPS distributions are treated as ground truth, they are still subject to randomness. If the total FPS sample size for a given player-game pair is insufficient—indicating that the player has not played the game long enough—the resulting FPS distribution may be distorted.

% For ground truth, multiple play records per player-game pair were aggregated to form an FPS distribution, using thresholds of 25Hz, 45Hz, 60Hz, and 145Hz to define gameplay smoothness, aligning with common standards to cover diverse game scenarios and reduce randomness.

% For ground truth, multiple play records for the same player-game pair were aggregated to derive an FPS distribution, increasing data points and covering diverse game scenarios to reduce randomness. We categorized FPS using thresholds commonly used in gaming: 24Hz, 45Hz, 60Hz, and 144Hz~\cite{hpframerate, infoframerate}. These thresholds define gameplay smoothness: below 24Hz causes noticeable stuttering, 24-45Hz is basic, 45-60Hz is stable, 60-144Hz is dynamic (suitable for e-sports), and above 144Hz offers exceptionally smooth gameplay. Due to the quantification during the data collection, the actual thresholds used are 25Hz, 45Hz, 60Hz, and 145Hz.

\subsection{Methodology}\label{methodology}

Let's revisit the problem definition: Given the characteristics of players and games, the goal is to predict the FPS performance of a new player-game pair. Each input data point consists of one-dimensional features, so we use the MLP (Multi-Layer Perceptron) as foundational blocks. 

Despite collecting as many quantifiable features as possible within ethical boundaries, each user and game has habits or characteristics that are hard to quantify and need case-by-case consideration. To address this problem, we assign each player and game a unique learnable knowledge kernel (LKK), distributed and trained on each player’s device. This ensures that these kernels accurately reflect their habits and characteristics, enhancing prediction accuracy and providing valuable customization.

% For example, consider two players with the same software and hardware. Player A’s device is used for both work and gaming, often running programs like Chrome and Microsoft PowerPoint in the background. In contrast, Player B’s device is dedicated to gaming, with all resources available during gameplay. Thus, Player B is likely to have higher runtime FPS. Similarly, Player C prefers running games with ray tracing function turned on, while Player D is content without that. Under identical conditions, Player C is likely to have lower FPS.

% This also applies to games. Each game requires specific learnable kernels for accurate FPS prediction. Even if two games are similar in genre, popularity, review score, release date, etc., their runtime FPS can differ significantly on the same device due to factors like optimization, default or recommended graphic settings, graphics finesse, texture detail, and scene complexity.

\begin{figure*}[]
	\includegraphics[width=\textwidth]{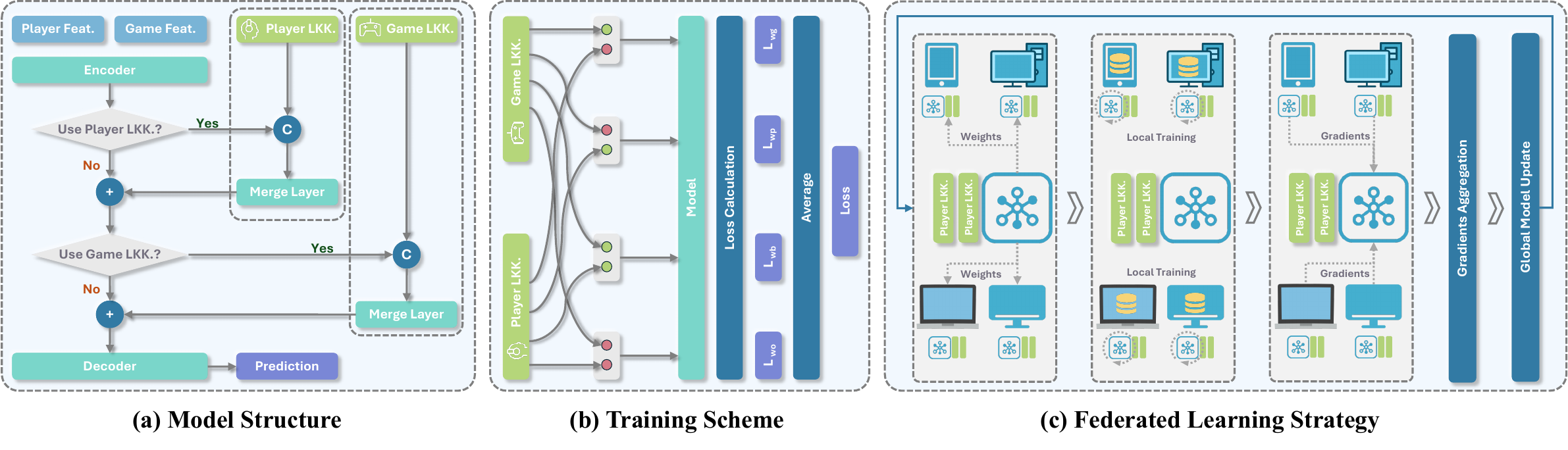}
	\caption{(a) illustrates the model structure and plug-and-play LKK usage. ``+'' in this figure means point-wise addition, and ``C'' means concatenate. (b) presents the workflow and loss calculation process, where red and green dots mean corresponding LKK ``not used'' and ``used'', respectively. (c) visualizes the federated learning scheme used in this work.}
	\label{fig:pipeline}
\end{figure*}

The cold start is also an issue. For newly released games or first-time users, the corresponding LKK isn't fully trained. To handle this, we propose a flexible approach to apply these kernels. This includes scenarios such as not using kernels, using only player or game-specific LKKs, and using both kernels. During training, we calculate the predicted FPS distribution and corresponding losses for each case separately and average them as the final loss (see Fig.~\ref{fig:pipeline}b). When applying an kernel, the main feature map is concatenated with the kernel and then fused through a Linear Merge layer with a skip connection. Therefore, the information that kernel brings refines the main feature map.

Denoting the encoder, decoder, merge layers for player and game LKKs as $E, D, M_p, M_g$, and input feature maps as $I$, we have:
\begin{align}
    O_{wo}&=D(E(I))\\
    O_{wp}&=D(E(I) + M_p(E(I)||K_p))\\
    O_{wg}&=D(E(I) + M_g(E(I)||K_g))\\
    O_{wb}&=D(E(I) + M_p(E(I)||K_p) + M_g(M_p(E(I)||K_p)||K_g))
\end{align}
, where $O_{wo}, O_{wp}, O_{wg}, O_{wb}$ means output under the condition of without LKK, with player LKK, with game LKK, with both LKK. The training loss can be further written as:
\begin{equation}
    L=\frac{1}{4}\sum_{k\in\mathcal{K}}{criterion(O_k, GT)}
\end{equation}
, where $\mathcal{K}$ represents the set consists of the four LKK usage cases mentioned above.

During testing, the conditions for applying LKKs can be preset, such as requiring at least three trained records for a player to enable the player-specific LKK, or ten records for a game to enable the game-specific LKK. When these conditions are met, the corresponding kernels are used in predictions (Fig.~\ref{fig:pipeline}a). The model's performance under the four cases is analyzed in Section \ref{sec:ablation}.

While this predictor benefits both players and developers, privacy remains a concern. Player-side features such as hardware specifications and geographic information are used as input, and the FPS distribution ground truth for purchased games depends on numerous FPS samples collected during gameplay. These samples can reveal sensitive details such as game session length and frequency.

The collection and transmission of user information to servers pose privacy risks. To mitigate this, we adopted a federated learning approach. Data are stored exclusively on users' devices, and models are trained locally. After local training, only gradients from client models are sent to the server, where they are aggregated to update the global model once gradients from a sufficient number of players are received. Updated model weights are then distributed to all client models again (see Fig.~\ref{fig:pipeline}c).

%% file: contents/6-results.tex
\section{Results of the FPS Predictor}
\begin{table*}[]
\centering
\caption{Results table of centeralized and fedrated learning models. WD: Wasserstein distance, CE: Cross Entropy, KL Div: Kullback–Leibler divergence, Adj Acc: Adjacent Accuracy. MAE: Mean Absolute Error.}
\begin{tabular}{lrrrrrrrr}
% \hline
\hlineB{2}
% \toprule
\cellcolor{gray!20}\textbf{Method} &  \cellcolor{gray!40}\textbf{WD $\downarrow$} & \cellcolor{gray!40}\textbf{CE$\downarrow$} & \cellcolor{gray!20}\textbf{MAE$\downarrow$} & \cellcolor{gray!20}\textbf{KL Div$\downarrow$} & \cellcolor{gray!20}\textbf{Top1 Acc$\uparrow$} & \cellcolor{gray!20}\textbf{Top2 Acc$\uparrow$} & \cellcolor{gray!20}\textbf{Adj Acc$\uparrow$} & \cellcolor{gray!20}\textbf{Top1 F1$\uparrow$} \\ 
\hlineB{1}
% \midrule
Softmax Regression               & 0.9713 & 1.6059 &  0.2179 &   0.8825  & 0.3774 &         0.6375 &        0.8189 &    0.2553    \\
Decision Tree                    & 0.5952 & 1.4570 &  0.1560 &  0.6414  & 0.5489 &         0.7983 &        0.9006 &    0.4042     \\
Random Forest                    & 0.5726 & 1.4510 &  0.1493 &   0.6110  & 0.5884 &         0.8148 &        0.9012 &    0.4102     \\
XGBoost \cite{chen2016xgboost}   & 0.6000 & 1.4716 &  0.1556 &   0.5377  & 0.5700 &         0.8012 &        0.8966 &    0.3896     \\
\hlineB{1}
% \midrule
Centralized                      & 0.4698 & 1.3935 &  \textbf{0.1302} &   \textbf{0.5188} & \textbf{0.6192} &         \textbf{0.8511} &        \textbf{0.9251} &    0.4677    \\ 
Federated                        & \textbf{0.4690} & \textbf{1.3871} &  \textbf{0.1302} &   0.5369 & \textbf{0.6192} &         0.8477 &        0.9221 &    \textbf{0.4741}   \\ 
\hlineB{2}
% \bottomrule
\end{tabular}
\label{tab:results}
\end{table*}
\subsection{5-Classes FPS Distribution Prediction}
To fully evaluate the performance of our proposed model and training strategy, we compared the results against several baseline methods and between centralized and federated training strategies. The results, shown in Table~\ref{tab:results}, demonstrate that our model outperforms all baseline methods across all metrics. Importantly, federated training does not degrade performance, as players' data remains local, and only gradients are shared with the server. These findings confirm that the federated training strategy is effective for this task, and our method is robust.

Our goal is to predict the distribution of FPS bins for a player-game pair. The primary metric we use is the Wasserstein distance, also known as Earth Mover's Distance (EMD), which is valuable for comparing probability distributions over discrete bins. It quantifies the minimum cost required to transform one distribution into another, based on the amount of probability mass moved and the distance it is moved. This metric is particularly suitable for this task, as it captures the differences between predicted and ground truth distributions while considering the bin structure. Mathematically, the Wasserstein distance of order one is defined as:
\begin{equation}
    W_1(P, Q) = \inf_{\gamma \in \Gamma(P, Q)} \int d(x, y) \, d\gamma(x, y)
\end{equation}
, where \( \Gamma(P, Q) \) denotes the set of all joint distributions \( \gamma \) whose marginals are \( P \) and \( Q \), and \( d(x, y) \) represents the distance between bins \( x \) and \( y \).
% _{X \times X}

Other commonly used metrics are also adopted to profile performance from various angles. These metrics include those focusing on distribution, such as cross entropy, Kullback–Leibler divergence, and mean absolute error, as well as category-centered metrics for profiling the predicted category with the highest probability, such as Top-1 accuracy and F1 score. Another metric, adjacent accuracy, measures the proportion of cases where the predicted top-1 FPS distribution category is the same as or adjacent to the ground truth category. This is used because the bin categories are not distinct classes, and we want to assess how close the predicted top-1 bins are to the ground truth.

\begin{figure}[h]
    \centering
    \begin{subfigure}[b]{0.48\textwidth}
        \includegraphics[width=\textwidth]{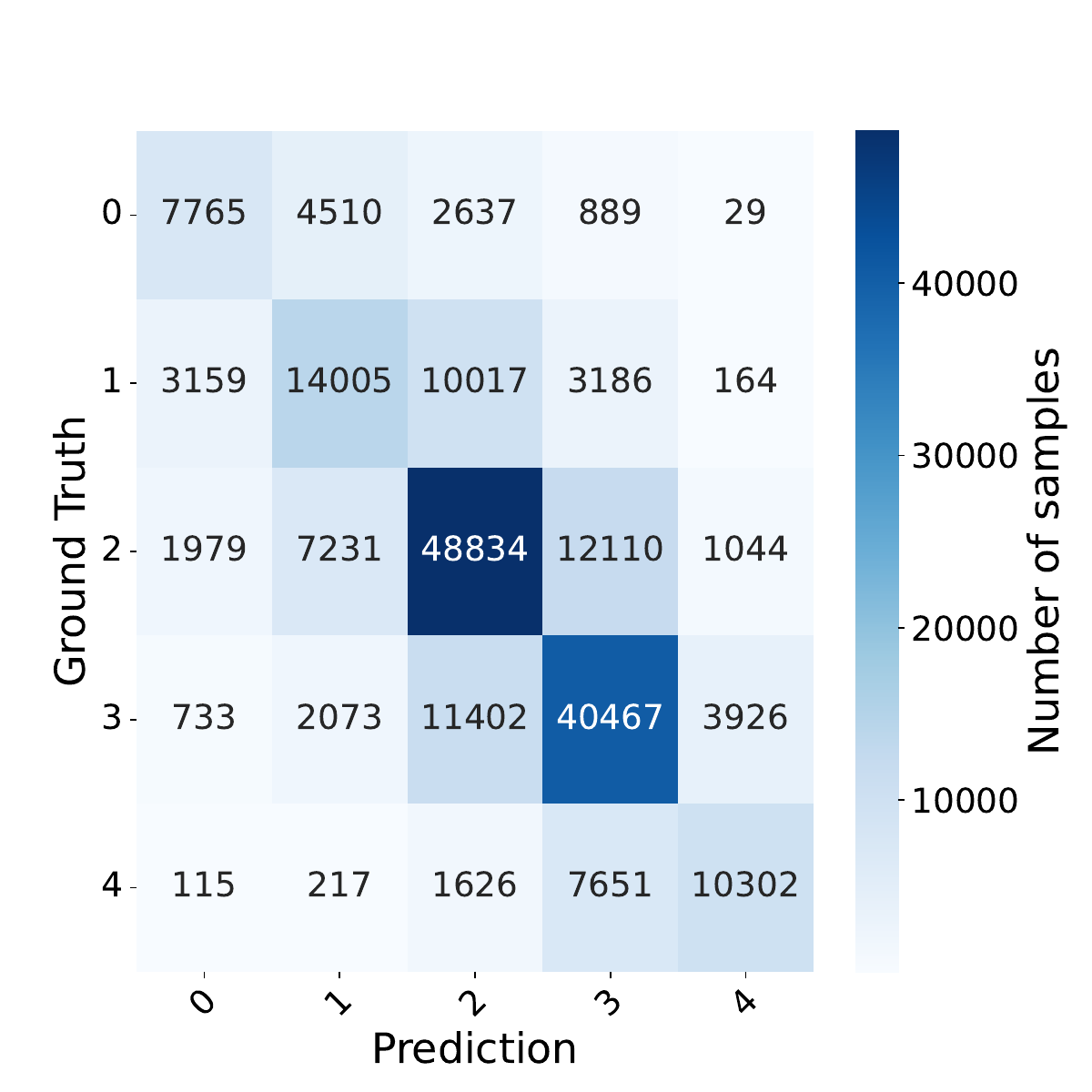}
        \caption{Centralized Training.}
        \label{fig:cent-cm}
    \end{subfigure}
    \hspace{0.02\textwidth}
    ~
    \begin{subfigure}[b]{0.48\textwidth}
        \includegraphics[width=\textwidth]{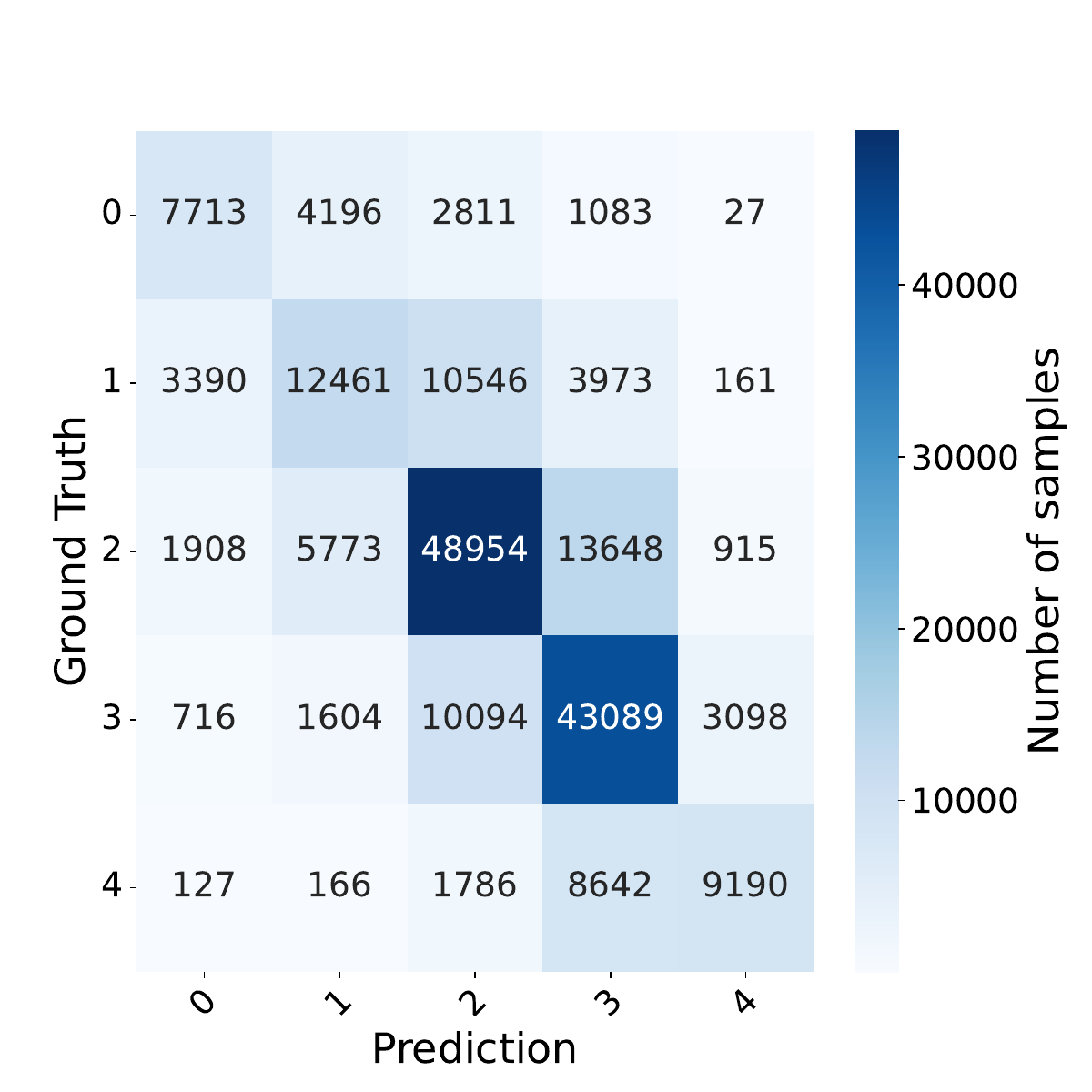}
        \caption{Federated Training.}
        \label{fig:fed-cm}
    \end{subfigure}
    \caption{Confusion matrices for top-1 FPS bin categories comparing centralized and federated FPS predictors. Predictions are concentrated along the diagonal, with the federated model showing performance similar to the centralized one.}
    \label{fig:fed-cent-cm}
\end{figure}

The confusion matrix for the best centralized and federated trained models is shown in Fig.~\ref{fig:fed-cm}. The matrix indicates that most predictions are concentrated on or near the diagonal, demonstrating that in most cases, our proposed model's predictions are either the same as or adjacent to the corresponding ground truth labels. Additionally, a comparison between Fig.~\ref{fig:fed-cm} and Fig.~\ref{fig:cent-cm} shows that the federated training strategy does not significantly impair the model's performance.

\begin{figure}[h]
\centering
\includegraphics[width=\textwidth]{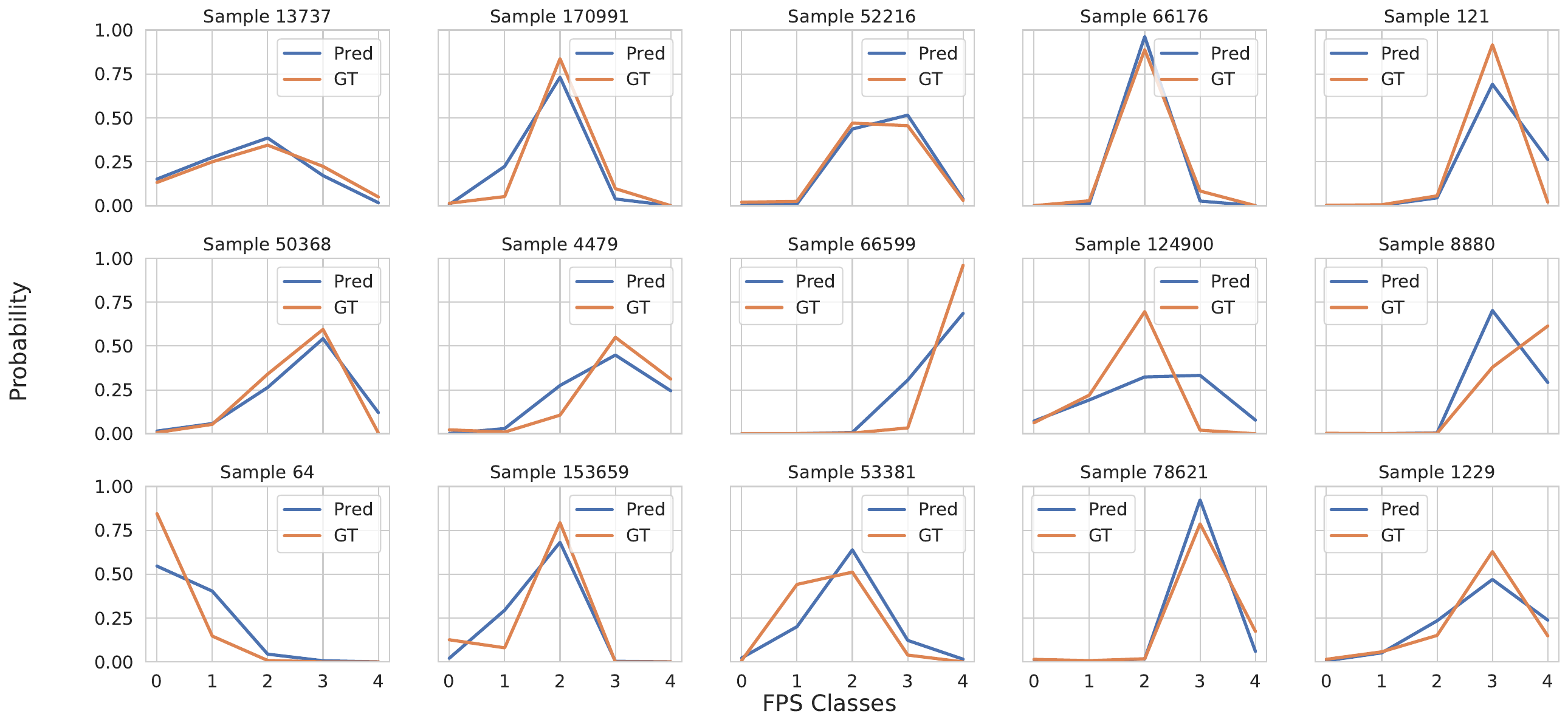}
\caption{Samples from the validation set, showing strong alignment between predicted and ground truth.}
\label{fig:results-samples}
\end{figure}

Fig.~\ref{fig:results-samples} provides a comparison between predictions and ground truth for some randomly selected validation samples. As shown in the figure, most of the time, the distribution of our predicted FPS performance is very close to the ground truth. Even if there is a slight deviation in the top probability category, the predicted distribution is still highly informative. In practical applications, alongside providing the most likely prediction, this approach can also offer users the predicted original distribution for reference if needed.

\subsection{Ablation Study}\label{sec:ablation}

LKK-related training and cold start strategies are core components of the proposed model. To evaluate the effectiveness of player and game-specific learnable knowledge kernels, we conducted an ablation study. We assessed the performance of the fully trained model on the validation set under four conditions: without any kernel, with only a game-specific learnable knowledge kernel, with only a player-specific learnable knowledge kernel, and with both types of kernels. The results, shown in Table~\ref{tab:ablation}, indicate that the best performance across all metrics is achieved when both kernels are used. Performance improves with the use of either kernel type compared to not using any knowledge kernel at all.

\begin{table}[ht]
\caption{Ablation study results. ``w/'' and ``w/o'' means ``with'' and ``without'' certain type of learnable knowledge kernel.}
% \begin{tabularx}{\textwidth}{XXXXX}%{lllll}
\begin{tabular}{lrrrrr}
\hlineB{2}
\cellcolor{gray!20}\textbf{Kernels} & \cellcolor{gray!40}\textbf{WD$\downarrow$} & \cellcolor{gray!40}\textbf{CE$\downarrow$} & \cellcolor{gray!20}\textbf{MAE$\downarrow$}  & \cellcolor{gray!20}\textbf{Acc$\uparrow$}  & \cellcolor{gray!20}\textbf{F1$\uparrow$} \\ 
% \hline
\hlineB{1}
w/o &     0.5074 &  1.4275 & 0.1375 & 0.6073 &    0.4337 \\
w/ player &   0.4781 &  1.3916 & 0.1325 & 0.6121 &    0.4696 \\
w/ game &     0.4984 &  1.4225 & 0.1354 & 0.6110 &    0.4387 \\
w/ both  &    \textbf{0.4690} & \textbf{1.3871} & \textbf{0.1302} &  \textbf{0.6192} &  \textbf{0.4741} \\
\hlineB{2}
\end{tabular}
\label{tab:ablation}
\end{table}

\subsection{Training Details}

The dataset was randomly split into 80\% training and 20\% validation subsets. We trained our neural network using the Adam optimizer (learning rate = 0.001) for 100 epochs, with L1 regularization (weight = \(1 \times 10^{-9}\)). Built on PyTorch, our model is 1.44 MB without embeddings, requiring 213,504 FLOPs for inference without embeddings and 377,344 FLOPs with embeddings. On an NVIDIA RTX 4090, the average inference time per sample is 0.14 ms with embeddings and 0.07 ms without. On a 13th Gen Intel(R) Core(TM) i9-13900 CPU, inference takes 1.15 ms with embeddings and 0.96 ms without. These results indicate that our model is lightweight and efficient across different hardware configurations.

\subsection{Auxillary Full 42-Classes FPS Distribution Prediction}

\begin{figure}[H]
\centering
\includegraphics[width=\textwidth]{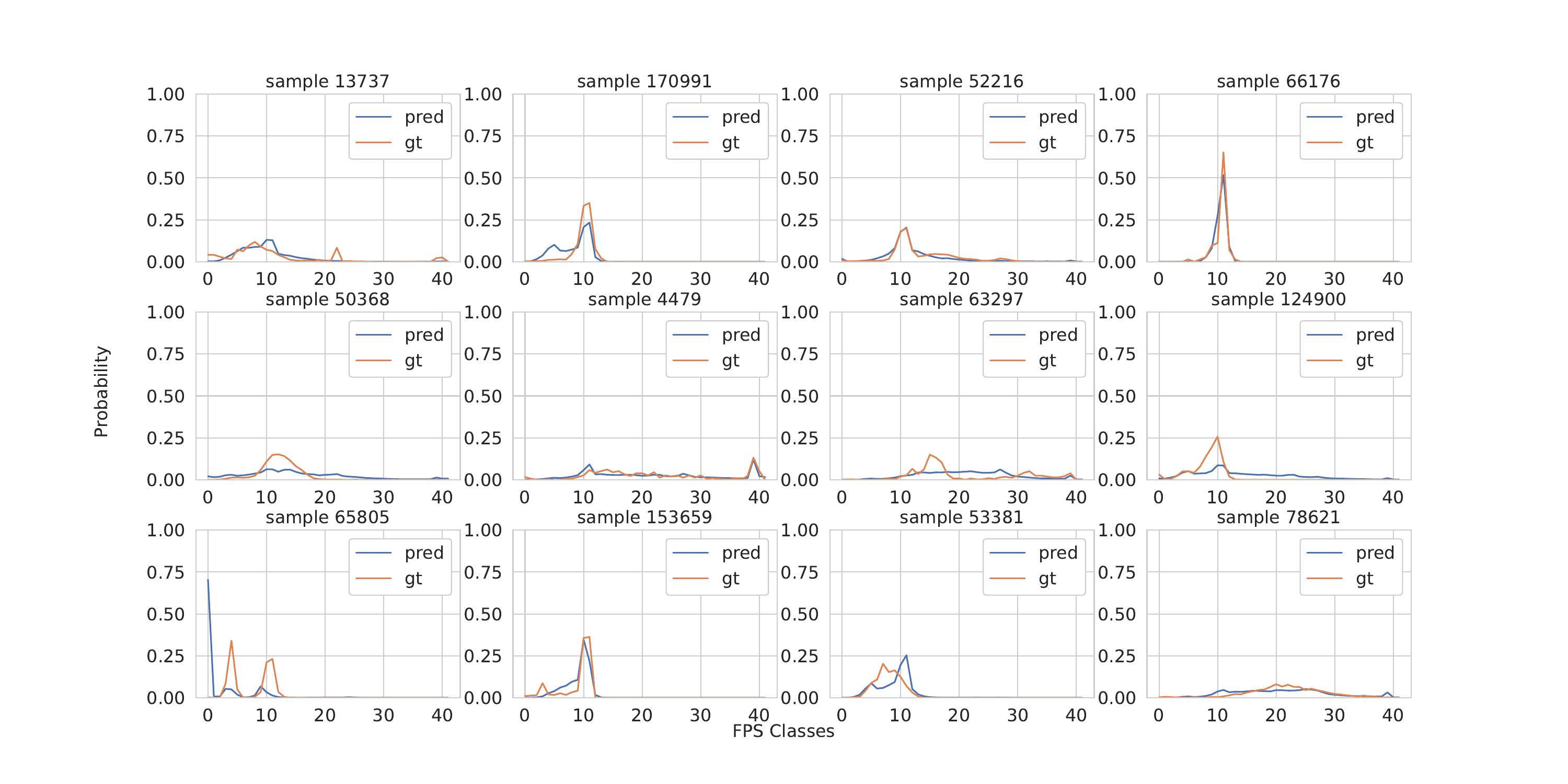}
\caption{Some randomly-selected sample results from the validation set on full 42-classes FPS distribution prediction.}
\label{fig:results-samples-42classes}
\end{figure}

For most players, the 5-class game FPS prediction is sufficient to help them make appropriate game purchase decisions while avoiding overwhelming yet unnecessary information. These five classes correspond to different levels of video game smoothness, providing better explainability. However, for professional players, a more fine-grained FPS distribution can offer better insights into the game's performance. For example, the concentration of the FPS distribution characterizes the stability of FPS performance, multiple local maxima in the FPS distribution may indicate significant differences in FPS performance across different in-game scenarios, and the proportion of low frame rates reflects the frequency of stuttering and negative gaming experiences. This information can help professionals make more well-informed decisions.

To address this need, we developed a 42-class FPS distribution predictor trained directly on the raw FPS distribution ground truth, without aggregating the outputs into coarser 5-class categories. This model prioritizes capturing the overall trend of the FPS distribution across classes, rather than maximizing accuracy in the top-1 prediction. Evaluation results include a distribution MAE of 0.0213, top-1 accuracy of 0.3473, and adjacent accuracy of 0.4808. These metrics demonstrate high distribution-level fidelity, indicating the robustness and reliability of our approach in capturing subtle variations in FPS trends. A detailed visualization of the 42-class prediction outcomes across sample cases is presented in Fig. \ref{fig:results-samples-42classes} (compared with the coarse 5-class results shown in Fig. \ref{fig:results-samples}). Together, these visualizations illustrate the model's capability to effectively capture both coarse and fine-grained trends in the 95\% FPS floor distribution, thereby showcasing the strength of our proposed model and training scheme in accurately predicting nuanced distribution patterns.

%% file: contents/7-discussion.tex
\section{Discussion and Future Work}

% 1. Discuss the significance of the world-scale data and the importance of the game experience modeling.

% 2. Briefly discuss why its such a challenging problem. Whst are the confounding factor and the importance of a privacy sensitive approach.

% 3. Discuss the novelty of our approach.

\subsection{Quantified Game Performance Analysis Based on the First Worldwide Gaming Telemetry Dataset}

The discussion on game performance has been ongoing since the inception of gaming. As games become more visually realistic and immersive, the computing power required to render them smoothly has increased significantly. The continuous advancement of game engines and gaming technology has driven parallel improvements in gaming hardware, optimizing performance and offering players a captivating audiovisual experience.

In the gaming community, runtime performance is recognized as a critical factor in the overall gaming experience, alongside quality and gameplay. Even with a well-crafted game concept and top-tier production, a laggy or inconsistent performance can render these qualities irrelevant, ultimately damaging a game's reputation, sales, and even the developer’s credibility.

While game performance is crucial, it is not the only factor affecting smooth gameplay. Network speed and server response times also play a role. However, in single-player and many online gaming scenarios, player-side device performance remains the predominant factor. To minimize performance-related issues, developers have made various efforts: testing games on diverse configurations to determine minimum and recommended specs, offering visual quality options to cater to different hardware capabilities, and releasing demo versions to let players gauge game performance firsthand. Recently, platforms like Xbox have begun utilizing deep learning techniques to predict a binary playability of games on individual devices, enhancing players' confidence in purchasing. Game console manufacturers also contribute by offering standardized configurations, enabling developers to optimize their games for specific hardware, which has played a significant role in the popularity of consoles worldwide.

% These include commonly cited elements such as GPU and CPU specifications, along with software configurations, geographical location, and numerous inherent game features. 

However, as previously discussed, game performance is influenced by a myriad of factors. Additionally, certain unquantifiable aspects—such as individual gaming habits, typical system settings, and the variations in optimization strategies across platforms—are challenging to capture on a per-game, per-user basis. Till now, analyzing these numerous influences on game performance holistically has been almost impossible. The industry-standard approach has been largely case-by-case, with developers providing minimum and recommended hardware specifications based on game-specific attributes, while players gauge potential performance based on their hardware and experience. This process is inefficient, inaccurate, and often leads to uncertainty and frustration, which can impact the game’s success.

The root of these challenges lies in the absence of a comprehensive, representative dataset that captures quantifiable factors influencing game performance. Additionally, there has been no approach capable of addressing unquantifiable aspects of performance in a way that satisfies both players and developers. Unlike previous efforts, our work introduces a large-scale dedicated dataset focused on this domain, capturing not only a wide range of user and game-specific characteristics but also providing robust ground-truth FPS distributions throughout complete game sessions. This unique dataset forms a reliable foundation for quantitative analysis and deep learning model training.

Our extensive, globally representative dataset highlights both shared and unique characteristics among players, providing a comprehensive opportunity to analyze game performance from multiple perspectives. Our proposed approach serves as a benchmark for game developers and game platforms, opening the door to greater transparency in game performance and benefiting the gaming community as a whole.

\subsection{Inherent Complexity and Entanglement of the Game Performance Prediction Task}
Predicting game performance presents a multifaceted challenge that arises from the necessity to integrate diverse data sources and account for numerous interdependent factors.

% \sssec{Design of the features to use.}

\sssec{Data Collection and Pre-processing.}
Many potential features could impact game performance. However, identifying the most promising ones is essential before beginning data collection. To achieve this, we conducted an extensive search of relevant materials, reviewed players' first-hand comments, and carried out trial-and-error analyses on smaller data samples. Data cleaning and pre-processing posed additional challenges, as some participants played games infrequently or not at all. Regional and language differences also introduced complications, as system and hardware labels were initially recorded in foreign languages, requiring remapping and filtering. Establishing standards for merging and filtering game records was crucial, as it directly influenced the quality of data used in downstream analysis and deep learning model training. Another major challenge is creating a mapping table to link game process names from various versions to unique official game names. Since there's no direct way to retrieve the official name of a running game, manually building a comprehensive mapping table is essential.

\sssec{Integration of Diverse Datasets.} Predicting game performance requires the integration of diverse datasets encompassing game-specific attributes, user characteristics, and hardware specifications. Currently, there are no comprehensive datasets that can cover all these aspects. Additionally, the heterogeneous nature of these data sources poses significant challenges in data merging, as there often lacks a common feature or identifier to serve as the bridge for integration. This fragmentation complicates the construction of a unified dataset, thereby hindering the development of accurate predictive models.

\sssec{Selection of Performance Metrics.} Many features influence overall game performance, making the choice of an effective metric to represent it both critical and challenging. Without a highly representative performance metric, accurately quantifying game runtime performance becomes difficult, hindering the development of a predictive model.

\sssec{Dynamic Nature of FPS.} FPS is a dynamic and fluctuating metric, influenced by real-time in-game activities and varying system conditions, which complicates prediction. The main challenge lies in deciding which FPS aspects to forecast—whether instantaneous FPS values, average FPS over specific intervals, or the distributional characteristics of the 95\% FPS floor performance. Addressing this requires transforming raw FPS data into a structured format that captures its temporal variability while preserving its value as a performance indicator.

% By modeling FPS distributions rather than single-point estimates, we can provide a more comprehensive and reliable prediction of game performance, effectively capturing the nuances of real-world gaming scenarios and ensuring that the predictions remain indicative of the actual user experience.

\sssec{The Cold Start Problem.} Another challenge in game performance prediction is the cold start problem, which occurs when the system encounters new users or games with limited historical data. For newly released games or first-time players, the lack of prior records hinders accurate predictions, affecting model generalization and reducing accuracy.  Addressing the cold start problem is essential for ensuring that the predictive system remains robust and functional across a diverse and evolving gaming landscape, where new games and players are continually introduced.

\sssec{Ensuring Data Privacy.} Data privacy always presents a significant obstacle in building prediction models involving user data due to the sensitive nature of the information required. Accurate FPS predictions often rely on detailed user-specific data, including hardware specifications, software configurations, and usage behaviors. Collecting and aggregating such data on centralized servers raises substantial privacy concerns, as it can expose users to risks related to data breaches and unauthorized access. Additionally, users may be hesitant to share personal information, leading to incomplete datasets that further complicate predictive accuracy. Balancing the need for comprehensive data with stringent privacy protections is a critical issue, necessitating the development of methodologies that safeguard user information while still enabling effective performance prediction.

% It requires data from games, user, and hardware. 1. usually don't have that much data. 2. data comes from different sources, how to merge those data together, usually there is no feature that can be used to do the merge

% there are multiple game related features, which one can best address game performance?

% fps is a constant changing feature and it's hard to do the prediction.

% how to solve the cold start problem?

% how to ensure data privacy

\begin{figure}[h]
\centering
\includegraphics[width=\textwidth]{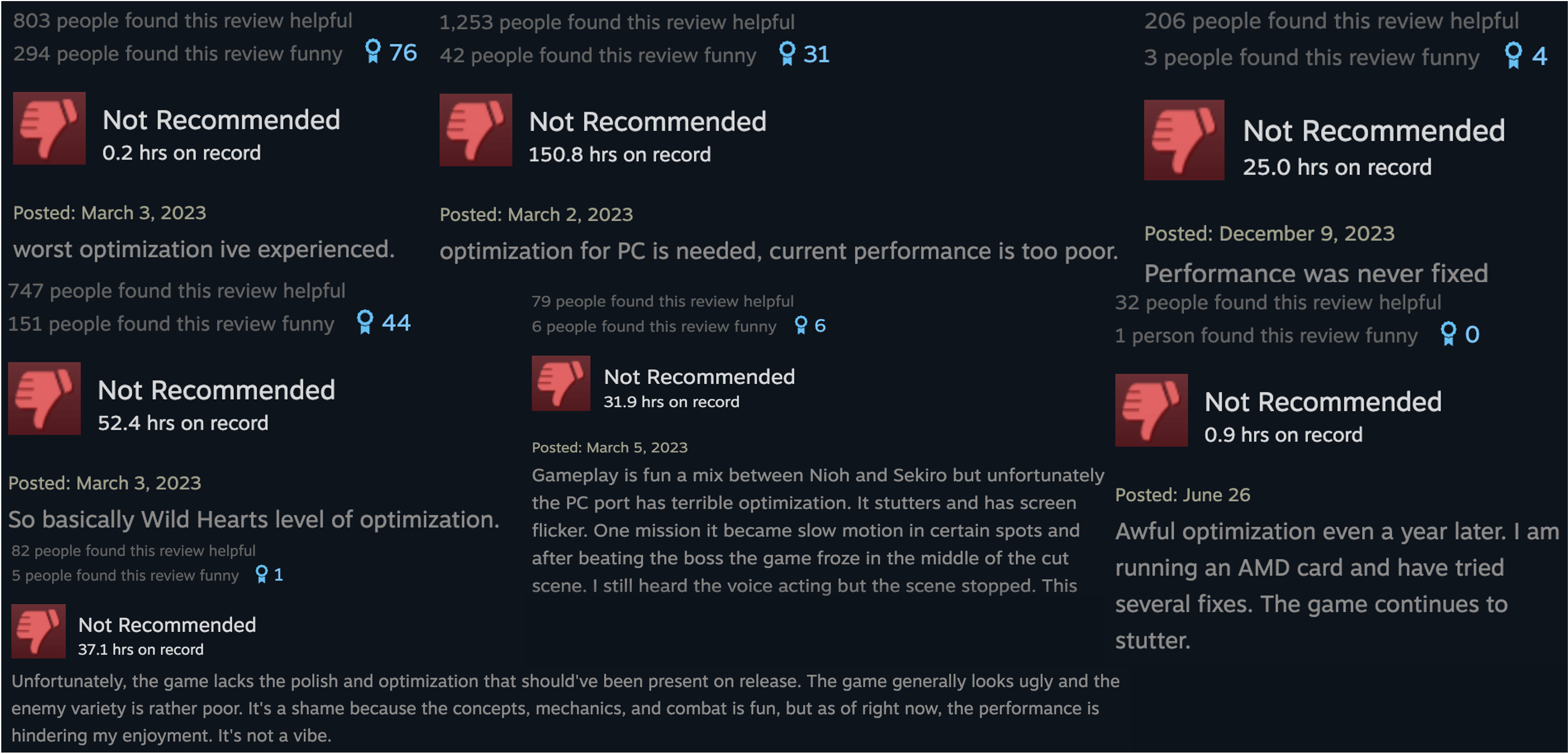}
\caption{Some examples of negative comments from Steam complaining about the poor game performance.}
\label{fig:negative-comments}
\end{figure} 

\subsection{Significance of the FPS Distribution Estimator}

This predictor is beneficial to players, game developers, and gaming platforms.

\sssec{For players: } FPS affects players' gaming performance and quality of experience. If the game cannot achieve the minimum tolerable FPS on a player's device, they are highly likely to hold a negative opinion of the game or give it up. Having an accurate FPS distribution estimation is essential to help players make wise and informed decisions, decreasing the likelihood of buying unsuitable games.

\sssec{For Game Developers: }
On platforms like Steam, poor game optimization and low FPS often lead to player dissatisfaction and commercial setbacks, as seen with titles like ``Cyberpunk 2077'' \cite{cyberpunk2077} and ``Nobunaga’s Ambition'' \cite{Nobunaga}. Some real negative feedbacks caused by poor game performance can be seen in Fig. \ref{fig:negative-comments}. While enhancing game quality and investing more in optimazation is ideal, letting potential customer's know about the estimated FPS performance can help with expectation management and mitigate negative reviews and revenue loss.

\sssec{For Gaming Platforms: }
Accurate FPS performance predictions can enhance platform service by better aligning game offerings with user hardware and habits. This proactive approach not only boosts customer satisfaction but also supports game developers by refining performance insights and consulting services.

In summary, integrating FPS performance predictors benefits players, developers, and platforms, aiming to improve the gaming experience and industry standards.

%% file: contents/ethical.tex
\section{Ethical Considerations}
Our institution's legal department reviewed the telemetry data collection process to ensure compliance with the California Consumer Privacy Act and European GDPR standards. We included only users who explicitly consented to participate voluntarily and ensured their anonymization throughout the data collection process. We do not collect any identifiable information, such as names, addresses, or genders. Users can withdraw from the project at any time and are not exposed to any harm from this data collection. The data are securely stored in an internal data center, with access restricted to those who sign a confidentiality agreement. Permission for the executable-to-game mapping table was obtained from the SteamDB administrator before web scraping, and data from IGDB were retrieved through its public API~\cite{IGDBAPI} in compliance with its regulations.

This dataset was collected under the guidance of the institution's confidentiality policy. Due to the extensive coverage and high sensitivity of the involved areas, it cannot be directly made available to the public. For academic or commercial purposes, interested parties may contact the designated company representative for potential access under confidentiality agreements. Following the principle of anonymity, the affiliated institution will only be disclosed once this article is published. The IGDB game-centric dataset and game-to-process name mapping table are available upon request.

%% file: contents/8-conclusion.tex
\section{Conclusion}
FPS is a crucial factor in determining the gaming experience. This paper designs a targeted large-scale data collection specifically for this factor, filling a gap in the quantitative analysis of big data. By analyzing software and hardware characteristics and game features from around the world, this paper delves deeply into various aspects affecting FPS from both micro and macro perspectives, providing well-founded analyses. Additionally, to address the issue of not knowing how a game will perform on a player's machine before purchase, this paper proposes a predictor based on federated learning, which takes into account the unique characteristics of both players and games. The predictor can provide results with various levels of granularity according to user needs, fully considering the cold start problem. This solution offers players more transparent and reliable information, boosting their purchasing confidence and providing insights into solving this type of issue.

%% file: main.bbl

\begin{thebibliography}{58}

%%% the Web address.

\ifx \showCODEN    \undefined \def \showCODEN     #1{\unskip}     \fi
\ifx \showDOI      \undefined \def \showDOI       #1{#1}\fi
\ifx \showISBNx    \undefined \def \showISBNx     #1{\unskip}     \fi
\ifx \showISBNxiii \undefined \def \showISBNxiii  #1{\unskip}     \fi
\ifx \showISSN     \undefined \def \showISSN      #1{\unskip}     \fi
\ifx \showLCCN     \undefined \def \showLCCN      #1{\unskip}     \fi
\ifx \shownote     \undefined \def \shownote      #1{#1}          \fi
\ifx \showarticletitle \undefined \def \showarticletitle #1{#1}   \fi
\ifx \showURL      \undefined \def \showURL       {\relax}        \fi
% invisible to TeX
\providecommand\bibfield[2]{#2}
\providecommand\bibinfo[2]{#2}
\providecommand\natexlab[1]{#1}
\providecommand\showeprint[2][]{arXiv:#2}

\bibitem[WoL(2023)]%
        {WoLongDev}
 \bibinfo{year}{2023}\natexlab{}.
\newblock \bibinfo{title}{Wo {{Long}} Dev Says Sorry about the Crappy {{PC}} Version but the Thing Is, You See, There Are Lots of Types of {{PC}} {\textbar} {{PC Gamer}}}.
\newblock \bibinfo{howpublished}{\url{https://www.pcgamer.com/wo-long-dev-says-sorry-about-the-crappy-pc-version-but-the-thing-is-you-see-there-are-lots-of-types-of-pc/}}.
\newblock


\bibitem[bla(2024)]%
        {blackmythwukong}
 \bibinfo{year}{2024}\natexlab{}.
\newblock \bibinfo{title}{Black Myth: Wukong on {{Steam}}}.
\newblock \bibinfo{howpublished}{\url{https://store.steampowered.com/app/2358720/Black_Myth_Wukong/}}.
\newblock


\bibitem[Epi(2024)]%
        {EpicGamesStore2024}
 \bibinfo{year}{2024}\natexlab{}.
\newblock \bibinfo{title}{Epic {{Games Store}}}.
\newblock \bibinfo{howpublished}{https://store.epicgames.com/en-US/}.
\newblock


\bibitem[Gra(2024)]%
        {GrandTheftAuto}
 \bibinfo{year}{2024}\natexlab{}.
\newblock \bibinfo{title}{Grand {{Theft Auto V}} -Configuration}.
\newblock \bibinfo{howpublished}{\url{https://steamdb.info/app/271590/config/}}.
\newblock


\bibitem[cyb(2024)]%
        {cyberpunk2077}
 \bibinfo{year}{2024}\natexlab{}.
\newblock \bibinfo{title}{{{Home of the Cyberpunk 2077 Universe --- Games, Anime \& More}}}.
\newblock \bibinfo{howpublished}{\url{https://www.cyberpunk.net/us/en/}}.
\newblock


\bibitem[IGD(2024a)]%
        {IGDBAPI}
 \bibinfo{year}{2024}\natexlab{a}.
\newblock \bibinfo{title}{{{IGDB}}: {{Video Game Database API}}}.
\newblock \bibinfo{howpublished}{\url{https://www.igdb.com/api}}.
\newblock


\bibitem[IGD(2024b)]%
        {IGDBComDiscover}
 \bibinfo{year}{2024}\natexlab{b}.
\newblock \bibinfo{title}{{{IGDB}}.Com - {{Discover}}, {{Rate}} \& {{Track Your Games}} {\textbar} {{Contribute}} to the {{Largest Video Games Database}}}.
\newblock \bibinfo{howpublished}{\url{https://www.igdb.com/}}.
\newblock


\bibitem[IMF(2024)]%
        {IMFData}
 \bibinfo{year}{2024}\natexlab{}.
\newblock \bibinfo{title}{{{IMF Data}}}.
\newblock \bibinfo{howpublished}{\url{https://www.imf.org/en/Data}}.
\newblock


\bibitem[Nob(2024)]%
        {Nobunaga}
 \bibinfo{year}{2024}\natexlab{}.
\newblock \bibinfo{title}{NOBUNAGA'S AMBITION on Steam}.
\newblock \bibinfo{howpublished}{\url{https://store.steampowered.com/app/544990/NOBUNAGAS_AMBITION/}}.
\newblock


\bibitem[Ste(2024a)]%
        {SteamStore}
 \bibinfo{year}{2024}\natexlab{a}.
\newblock \bibinfo{title}{Steam {{Store}}}.
\newblock \bibinfo{howpublished}{\url{https://store.steampowered.com/}}.
\newblock


\bibitem[Ste(2024b)]%
        {SteamDB}
 \bibinfo{year}{2024}\natexlab{b}.
\newblock \bibinfo{title}{{{SteamDB}}}.
\newblock \bibinfo{howpublished}{\url{https://steamdb.info/}}.
\newblock


\bibitem[Dir(2024)]%
        {DirectX}
 \bibinfo{year}{2024}\natexlab{}.
\newblock \bibinfo{title}{{What is DirectX?} - {{Microsoft Support}}}.
\newblock \bibinfo{howpublished}{\url{https://support.microsoft.com/en-us/topic/how-to-install-the-latest-version-of-directx-d1f5ffa5-dae2-246c-91b1-ee1e973ed8c2}}.
\newblock


\bibitem[Why(2024)]%
        {WhyCyberpunk2077}
 \bibinfo{year}{2024}\natexlab{}.
\newblock \bibinfo{title}{Why Is {{Cyberpunk}} 2077 so Horribly Optimised for {{PC}}, despite All the Delays and Promises?}
\newblock \bibinfo{howpublished}{\url{https://www.quora.com/Why-is-Cyberpunk-2077-so-horribly-optimised-for-PC-despite-all-the-delays-and-promises}}.
\newblock


\bibitem[Wor(2024)]%
        {WorldBank}
 \bibinfo{year}{2024}\natexlab{}.
\newblock \bibinfo{title}{World {{Bank Open Data}}}.
\newblock \bibinfo{howpublished}{\url{https://data.worldbank.org}}.
\newblock


\bibitem[Xbo(2024)]%
        {XboxGames}
 \bibinfo{year}{2024}\natexlab{}.
\newblock \bibinfo{title}{Xbox {{Games}}}.
\newblock \bibinfo{howpublished}{\url{https://www.xbox.com/en-US/games}}.
\newblock


\bibitem[Awosika et~al\mbox{.}(2024)]%
        {awosikaTransparencyPrivacyRole2024}
\bibfield{author}{\bibinfo{person}{Tomisin Awosika}, \bibinfo{person}{Raj~Mani Shukla}, {and} \bibinfo{person}{Bernardi Pranggono}.} \bibinfo{year}{2024}\natexlab{}.
\newblock \showarticletitle{Transparency and {{Privacy}}: {{The Role}} of {{Explainable AI}} and {{Federated Learning}} in {{Financial Fraud Detection}}}.
\newblock \bibinfo{journal}{\emph{IEEE Access}}  \bibinfo{volume}{12} (\bibinfo{year}{2024}), \bibinfo{pages}{64551--64560}.
\newblock
\showISSN{2169-3536}


\bibitem[C et~al\mbox{.}(2022)]%
        {cFederatedLearningSmart2022}
\bibfield{author}{\bibinfo{person}{NguyenDinh C}, \bibinfo{person}{{PhamQuoc-Viet}}, \bibinfo{person}{PathiranaPubudu N}, \bibinfo{person}{DingMing}, \bibinfo{person}{SeneviratneAruna}, \bibinfo{person}{LinZihuai}, \bibinfo{person}{DobreOctavia}, {and} \bibinfo{person}{{HwangWon-Joo}}.} \bibinfo{year}{2022}\natexlab{}.
\newblock \showarticletitle{Federated {{Learning}} for {{Smart Healthcare}}: {{A Survey}}}.
\newblock \bibinfo{journal}{\emph{ACM Computing Surveys (CSUR)}} (\bibinfo{date}{Feb.} \bibinfo{year}{2022}).
\newblock


\bibitem[{Calvillo-G{\'a}mez} et~al\mbox{.}(2015)]%
        {calvillo-gamezAssessingCoreElements2015}
\bibfield{author}{\bibinfo{person}{Eduardo~H. {Calvillo-G{\'a}mez}}, \bibinfo{person}{Paul Cairns}, {and} \bibinfo{person}{Anna~L. Cox}.} \bibinfo{year}{2015}\natexlab{}.
\newblock \showarticletitle{Assessing the {{Core Elements}} of the {{Gaming Experience}}}.
\newblock In \bibinfo{booktitle}{\emph{Game {{User Experience Evaluation}}}}, \bibfield{editor}{\bibinfo{person}{Regina Bernhaupt}} (Ed.). \bibinfo{publisher}{Springer International Publishing}, \bibinfo{address}{Cham}, \bibinfo{pages}{37--62}.
\newblock
\showISBNx{978-3-319-15985-0}


\bibitem[Chatterjee et~al\mbox{.}(2024)]%
        {chatterjeeFederatedLearningEmpowered2024}
\bibfield{author}{\bibinfo{person}{Pushpita Chatterjee}, \bibinfo{person}{Debashis Das}, {and} \bibinfo{person}{Danda~B. Rawat}.} \bibinfo{year}{2024}\natexlab{}.
\newblock \showarticletitle{Federated {{Learning Empowered Recommendation Model}} for {{Financial Consumer Services}}}.
\newblock \bibinfo{journal}{\emph{IEEE Transactions on Consumer Electronics}} \bibinfo{volume}{70}, \bibinfo{number}{1} (\bibinfo{date}{Feb.} \bibinfo{year}{2024}), \bibinfo{pages}{2508--2516}.
\newblock
\showISSN{1558-4127}


\bibitem[Chen et~al\mbox{.}(2023)]%
        {chenFSREALRealWorldCrossDevice2023}
\bibfield{author}{\bibinfo{person}{Daoyuan Chen}, \bibinfo{person}{Dawei Gao}, \bibinfo{person}{Yuexiang Xie}, \bibinfo{person}{Xuchen Pan}, \bibinfo{person}{Zitao Li}, \bibinfo{person}{Yaliang Li}, \bibinfo{person}{Bolin Ding}, {and} \bibinfo{person}{Jingren Zhou}.} \bibinfo{year}{2023}\natexlab{}.
\newblock \showarticletitle{{{FS-REAL}}: {{Towards Real-World Cross-Device Federated Learning}}}. In \bibinfo{booktitle}{\emph{Proceedings of the 29th {{ACM SIGKDD Conference}} on {{Knowledge Discovery}} and {{Data Mining}}}} \emph{(\bibinfo{series}{{{KDD}} '23})}. \bibinfo{publisher}{Association for Computing Machinery}, \bibinfo{address}{New York, NY, USA}, \bibinfo{pages}{3829--3841}.
\newblock
\showISBNx{9798400701030}


\bibitem[Chen and Guestrin(2016)]%
        {chen2016xgboost}
\bibfield{author}{\bibinfo{person}{Tianqi Chen} {and} \bibinfo{person}{Carlos Guestrin}.} \bibinfo{year}{2016}\natexlab{}.
\newblock \showarticletitle{Xgboost: A scalable tree boosting system}. In \bibinfo{booktitle}{\emph{Proceedings of the 22nd acm sigkdd international conference on knowledge discovery and data mining}}. \bibinfo{pages}{785--794}.
\newblock


\bibitem[Claypool and Claypool(2007)]%
        {claypool2007frame}
\bibfield{author}{\bibinfo{person}{Kajal~T Claypool} {and} \bibinfo{person}{Mark Claypool}.} \bibinfo{year}{2007}\natexlab{}.
\newblock \showarticletitle{On frame rate and player performance in first person shooter games}.
\newblock \bibinfo{journal}{\emph{Multimedia systems}} \bibinfo{volume}{13}, \bibinfo{number}{1} (\bibinfo{year}{2007}), \bibinfo{pages}{3--17}.
\newblock


\bibitem[Claypool and Claypool(2009)]%
        {claypoolPerspectivesFrameRates2009}
\bibfield{author}{\bibinfo{person}{Mark Claypool} {and} \bibinfo{person}{Kajal Claypool}.} \bibinfo{year}{2009}\natexlab{}.
\newblock \showarticletitle{Perspectives, Frame Rates and Resolutions: It's All in the Game}. In \bibinfo{booktitle}{\emph{Proceedings of the 4th {{International Conference}} on {{Foundations}} of {{Digital Games}}}} \emph{(\bibinfo{series}{{{FDG}} '09})}. \bibinfo{publisher}{Association for Computing Machinery}, \bibinfo{address}{New York, NY, USA}, \bibinfo{pages}{42--49}.
\newblock
\showISBNx{978-1-60558-437-9}


\bibitem[Claypool et~al\mbox{.}(2006)]%
        {claypoolEffectsFrameRate2006}
\bibfield{author}{\bibinfo{person}{Mark Claypool}, \bibinfo{person}{Kajal Claypool}, {and} \bibinfo{person}{Feissal Damaa}.} \bibinfo{year}{2006}\natexlab{}.
\newblock \showarticletitle{The effects of frame rate and resolution on users playing first person shooter games}. In \bibinfo{booktitle}{\emph{Multimedia computing and networking 2006}}, Vol.~\bibinfo{volume}{6071}. SPIE, \bibinfo{pages}{607101}.
\newblock


\bibitem[Elbir and Coleri(2020)]%
        {elbirFederated2020}
\bibfield{author}{\bibinfo{person}{Ahmet~M. Elbir} {and} \bibinfo{person}{Sinem Coleri}.} \bibinfo{year}{2020}\natexlab{}.
\newblock \showarticletitle{Federated Learning for Hybrid Beamforming in mm-Wave Massive MIMO}.
\newblock \bibinfo{journal}{\emph{IEEE Communications Letters}} \bibinfo{volume}{24}, \bibinfo{number}{12} (\bibinfo{year}{2020}), \bibinfo{pages}{2795--2799}.
\newblock


\bibitem[Feng et~al\mbox{.}(2020)]%
        {10.1145/3381006}
\bibfield{author}{\bibinfo{person}{Jie Feng}, \bibinfo{person}{Can Rong}, \bibinfo{person}{Funing Sun}, \bibinfo{person}{Diansheng Guo}, {and} \bibinfo{person}{Yong Li}.} \bibinfo{year}{2020}\natexlab{}.
\newblock \showarticletitle{PMF: A Privacy-preserving Human Mobility Prediction Framework via Federated Learning}.
\newblock \bibinfo{journal}{\emph{Proc. ACM Interact. Mob. Wearable Ubiquitous Technol.}} \bibinfo{volume}{4}, \bibinfo{number}{1}, Article \bibinfo{articleno}{10} (\bibinfo{date}{March} \bibinfo{year}{2020}), \bibinfo{numpages}{21}~pages.
\newblock
\urldef\tempurl%
\url{https://doi.org/10.1145/3381006}
\showDOI{\tempurl}


\bibitem[Hefner et~al\mbox{.}(2007)]%
        {hefner2007identification}
\bibfield{author}{\bibinfo{person}{Doroth{\'e}e Hefner}, \bibinfo{person}{Christoph Klimmt}, {and} \bibinfo{person}{Peter Vorderer}.} \bibinfo{year}{2007}\natexlab{}.
\newblock \showarticletitle{Identification with the player character as determinant of video game enjoyment}. In \bibinfo{booktitle}{\emph{International conference on entertainment computing}}. Springer, \bibinfo{pages}{39--48}.
\newblock


\bibitem[Hinum(2024)]%
        {hinumMobileGraphicsCards}
\bibfield{author}{\bibinfo{person}{Klaus Hinum}.} \bibinfo{year}{2024}\natexlab{}.
\newblock \bibinfo{title}{Mobile {{Graphics Cards}} - {{Benchmark List}}}.
\newblock \bibinfo{howpublished}{https://www.notebookcheck.net/Mobile-Graphics-Cards-Benchmark-List.844.0.html}.
\newblock


\bibitem[IssaWael et~al\mbox{.}(2023)]%
        {issawaelBlockchainBasedFederatedLearning2023}
\bibfield{author}{\bibinfo{person}{IssaWael}, \bibinfo{person}{MoustafaNour}, \bibinfo{person}{TurnbullBenjamin}, \bibinfo{person}{SohrabiNasrin}, {and} \bibinfo{person}{TariZahir}.} \bibinfo{year}{2023}\natexlab{}.
\newblock \showarticletitle{Blockchain-{{Based Federated Learning}} for {{Securing Internet}} of {{Things}}: {{A Comprehensive Survey}}}.
\newblock \bibinfo{journal}{\emph{Comput. Surveys}} (\bibinfo{date}{Jan.} \bibinfo{year}{2023}).
\newblock


\bibitem[Jagoda and McDonald(2018)]%
        {jagoda2018game}
\bibfield{author}{\bibinfo{person}{Patrick Jagoda} {and} \bibinfo{person}{Peter McDonald}.} \bibinfo{year}{2018}\natexlab{}.
\newblock \showarticletitle{Game Mechanics, Experience Design, and Affective Play}.
\newblock In \bibinfo{booktitle}{\emph{The Routledge Companion to Media Studies and Digital Humanities}}, \bibfield{editor}{\bibinfo{person}{Jentery Sayers}} (Ed.). \bibinfo{publisher}{Routledge}, \bibinfo{address}{New York}, \bibinfo{pages}{174--182}.
\newblock


\bibitem[Karimireddy et~al\mbox{.}(2020)]%
        {karimireddySCAFFOLDStochasticControlled2020}
\bibfield{author}{\bibinfo{person}{Sai~Praneeth Karimireddy}, \bibinfo{person}{Satyen Kale}, \bibinfo{person}{Mehryar Mohri}, \bibinfo{person}{Sashank Reddi}, \bibinfo{person}{Sebastian Stich}, {and} \bibinfo{person}{Ananda~Theertha Suresh}.} \bibinfo{year}{2020}\natexlab{}.
\newblock \showarticletitle{{{SCAFFOLD}}: {{Stochastic Controlled Averaging}} for {{Federated Learning}}}. In \bibinfo{booktitle}{\emph{Proceedings of the 37th {{International Conference}} on {{Machine Learning}}}}. \bibinfo{publisher}{PMLR}, \bibinfo{pages}{5132--5143}.
\newblock
\showISSN{2640-3498}


\bibitem[Kashcha et~al\mbox{.}(2022)]%
        {kashcha2022country}
\bibfield{author}{\bibinfo{person}{Mariia Kashcha}, \bibinfo{person}{Valerii Yatsenko}, {and} \bibinfo{person}{Tam{\'a}s Gy{\"o}m{\"o}rei}.} \bibinfo{year}{2022}\natexlab{}.
\newblock \showarticletitle{Country performance in e-sport: Social and economic development determinants.}
\newblock \bibinfo{journal}{\emph{Journal of International Studies (2071-8330)}} \bibinfo{volume}{15}, \bibinfo{number}{4} (\bibinfo{year}{2022}).
\newblock


\bibitem[Klimmt et~al\mbox{.}(2019)]%
        {klimmt2019effects}
\bibfield{author}{\bibinfo{person}{Christoph Klimmt}, \bibinfo{person}{Daniel Possler}, \bibinfo{person}{Nicolas May}, \bibinfo{person}{Hendrik Auge}, \bibinfo{person}{Louisa Wanjek}, {and} \bibinfo{person}{Anna-Lena Wolf}.} \bibinfo{year}{2019}\natexlab{}.
\newblock \showarticletitle{Effects of soundtrack music on the video game experience}.
\newblock \bibinfo{journal}{\emph{Media Psychology}} \bibinfo{volume}{22}, \bibinfo{number}{5} (\bibinfo{year}{2019}), \bibinfo{pages}{689--713}.
\newblock


\bibitem[Kone{\v c}n{\'y} et~al\mbox{.}(2017)]%
        {konecnyFederatedLearningStrategies2017}
\bibfield{author}{\bibinfo{person}{Jakub Kone{\v c}n{\'y}}, \bibinfo{person}{H.~Brendan McMahan}, \bibinfo{person}{Felix~X. Yu}, \bibinfo{person}{Peter Richt{\'a}rik}, \bibinfo{person}{Ananda~Theertha Suresh}, {and} \bibinfo{person}{Dave Bacon}.} \bibinfo{year}{2017}\natexlab{}.
\newblock \bibinfo{title}{Federated {{Learning}}: {{Strategies}} for {{Improving Communication Efficiency}}}.
\newblock
\newblock
\showeprint[arxiv]{1610.05492}~[cs]


\bibitem[Li et~al\mbox{.}(2020)]%
        {liReviewApplicationsFederated2020}
\bibfield{author}{\bibinfo{person}{Li Li}, \bibinfo{person}{Yuxi Fan}, \bibinfo{person}{Mike Tse}, {and} \bibinfo{person}{Kuo-Yi Lin}.} \bibinfo{year}{2020}\natexlab{}.
\newblock \showarticletitle{A Review of Applications in Federated Learning}.
\newblock \bibinfo{journal}{\emph{Computers \& Industrial Engineering}}  \bibinfo{volume}{149} (\bibinfo{date}{Nov.} \bibinfo{year}{2020}), \bibinfo{pages}{106854}.
\newblock
\showISSN{0360-8352}


\bibitem[Lika et~al\mbox{.}(2014)]%
        {lika2014facing}
\bibfield{author}{\bibinfo{person}{Blerina Lika}, \bibinfo{person}{Kostas Kolomvatsos}, {and} \bibinfo{person}{Stathes Hadjiefthymiades}.} \bibinfo{year}{2014}\natexlab{}.
\newblock \showarticletitle{Facing the cold start problem in recommender systems}.
\newblock \bibinfo{journal}{\emph{Expert systems with applications}} \bibinfo{volume}{41}, \bibinfo{number}{4} (\bibinfo{year}{2014}), \bibinfo{pages}{2065--2073}.
\newblock


\bibitem[Lipscomb and Zehnder(2004)]%
        {lipscomb2004immersion}
\bibfield{author}{\bibinfo{person}{Scott~D Lipscomb} {and} \bibinfo{person}{Sean~M Zehnder}.} \bibinfo{year}{2004}\natexlab{}.
\newblock \showarticletitle{Immersion in the virtual environment: The effect of a musical score on the video gaming experience}.
\newblock \bibinfo{journal}{\emph{Journal of Physiological Anthropology and Applied Human Science}} \bibinfo{volume}{23}, \bibinfo{number}{6} (\bibinfo{year}{2004}), \bibinfo{pages}{337--343}.
\newblock


\bibitem[Liu et~al\mbox{.}(2023)]%
        {liuEffectsFrameRate2023}
\bibfield{author}{\bibinfo{person}{Shengmei Liu}, \bibinfo{person}{Atsuo Kuwahara}, \bibinfo{person}{James~J Scovell}, {and} \bibinfo{person}{Mark Claypool}.} \bibinfo{year}{2023}\natexlab{}.
\newblock \showarticletitle{The {{Effects}} of {{Frame Rate Variation}} on {{Game Player Quality}} of {{Experience}}}. In \bibinfo{booktitle}{\emph{Proceedings of the 2023 {{CHI Conference}} on {{Human Factors}} in {{Computing Systems}}}} \emph{(\bibinfo{series}{{{CHI}} '23})}. \bibinfo{publisher}{Association for Computing Machinery}, \bibinfo{address}{New York, NY, USA}, \bibinfo{pages}{1--10}.
\newblock
\showISBNx{978-1-4503-9421-5}


\bibitem[Mammen(2021)]%
        {mammenFederatedLearningOpportunities2021}
\bibfield{author}{\bibinfo{person}{Priyanka~Mary Mammen}.} \bibinfo{year}{2021}\natexlab{}.
\newblock \bibinfo{title}{Federated {{Learning}}: {{Opportunities}} and {{Challenges}}}.
\newblock
\newblock
\showeprint[arxiv]{2101.05428}~[cs]


\bibitem[McCarthy and Wright(2004)]%
        {mccarthy2004technology}
\bibfield{author}{\bibinfo{person}{John McCarthy} {and} \bibinfo{person}{Peter Wright}.} \bibinfo{year}{2004}\natexlab{}.
\newblock \showarticletitle{Technology as experience}.
\newblock \bibinfo{journal}{\emph{interactions}} \bibinfo{volume}{11}, \bibinfo{number}{5} (\bibinfo{year}{2004}), \bibinfo{pages}{42--43}.
\newblock


\bibitem[McEwan et~al\mbox{.}(2012)]%
        {mcewan2012videogame}
\bibfield{author}{\bibinfo{person}{Mitchell McEwan}, \bibinfo{person}{Daniel Johnson}, \bibinfo{person}{Peta Wyeth}, {and} \bibinfo{person}{Alethea Blackler}.} \bibinfo{year}{2012}\natexlab{}.
\newblock \showarticletitle{Videogame control device impact on the play experience}. In \bibinfo{booktitle}{\emph{Proceedings of the 8th australasian conference on interactive entertainment: Playing the system}}. \bibinfo{pages}{1--3}.
\newblock


\bibitem[Moll et~al\mbox{.}(2020)]%
        {moll2020players}
\bibfield{author}{\bibinfo{person}{Philipp Moll}, \bibinfo{person}{Veit Frick}, \bibinfo{person}{Natascha Rauscher}, {and} \bibinfo{person}{Mathias Lux}.} \bibinfo{year}{2020}\natexlab{}.
\newblock \showarticletitle{How players play games: observing the influences of game mechanics}. In \bibinfo{booktitle}{\emph{Proceedings of the 12th acm international workshop on immersive mixed and virtual environment systems}}. \bibinfo{pages}{7--12}.
\newblock


\bibitem[{National Institute of Standards and Technology / U.S. Department of Commerce}(2012)]%
        {TukeyMethod}
\bibfield{author}{\bibinfo{person}{{National Institute of Standards and Technology / U.S. Department of Commerce}}.} \bibinfo{year}{2012}\natexlab{}.
\newblock \showarticletitle{{Tukey's Method}}.
\newblock In \bibinfo{booktitle}{\emph{{NIST/SEMATECH e-Handbook of Statistical Methods}}}. Chapter 4.7.1.
\newblock
\urldef\tempurl%
\url{https://www.itl.nist.gov/div898/handbook/prc/section4/prc471.htm}
\showURL{%
\tempurl}
\newblock
\shownote{Accessed: 2024-08-01}.


\bibitem[Parshakov and Zavertiaeva(2018)]%
        {parshakov2018determinants}
\bibfield{author}{\bibinfo{person}{Petr Parshakov} {and} \bibinfo{person}{Marina Zavertiaeva}.} \bibinfo{year}{2018}\natexlab{}.
\newblock \showarticletitle{Determinants of performance in eSports: A country-level analysis}.
\newblock \bibinfo{journal}{\emph{International Journal of Sport Finance}} \bibinfo{volume}{13}, \bibinfo{number}{1} (\bibinfo{year}{2018}), \bibinfo{pages}{34--51}.
\newblock


\bibitem[PfeifferKilian et~al\mbox{.}(2023)]%
        {pfeifferkilianFederatedLearningComputationally2023}
\bibfield{author}{\bibinfo{person}{PfeifferKilian}, \bibinfo{person}{RappMartin}, \bibinfo{person}{KhaliliRamin}, {and} \bibinfo{person}{HenkelJ{\"o}rg}.} \bibinfo{year}{2023}\natexlab{}.
\newblock \showarticletitle{Federated {{Learning}} for {{Computationally Constrained Heterogeneous Devices}}: {{A Survey}}}.
\newblock \bibinfo{journal}{\emph{Comput. Surveys}} (\bibinfo{date}{July} \bibinfo{year}{2023}).
\newblock


\bibitem[Popper(2013)]%
        {popperKnowledgeBodyMindProblem2013}
\bibfield{author}{\bibinfo{person}{Karl Popper}.} \bibinfo{year}{2013}\natexlab{}.
\newblock \bibinfo{booktitle}{\emph{Knowledge and the {{Body-Mind Problem}}: {{In Defence}} of {{Interaction}}}}.
\newblock \bibinfo{publisher}{Routledge}, \bibinfo{address}{New York}.
\newblock
\showISBNx{978-1-135-97536-4}


\bibitem[Quirk(2012)]%
        {Quirk2012oneway}
\bibfield{author}{\bibinfo{person}{Thomas~J. Quirk}.} \bibinfo{year}{2012}\natexlab{}.
\newblock \bibinfo{booktitle}{\emph{One-Way Analysis of Variance (ANOVA)}}.
\newblock \bibinfo{publisher}{Springer New York}, \bibinfo{address}{New York, NY}, \bibinfo{pages}{163--179}.
\newblock
\showISBNx{978-1-4614-3725-3}
\urldef\tempurl%
\url{https://doi.org/10.1007/978-1-4614-3725-3_8}
\showDOI{\tempurl}


\bibitem[Rieke et~al\mbox{.}(2020)]%
        {riekeFutureDigitalHealth2020}
\bibfield{author}{\bibinfo{person}{Nicola Rieke}, \bibinfo{person}{Jonny Hancox}, \bibinfo{person}{Wenqi Li}, \bibinfo{person}{Fausto Milletar{\`i}}, \bibinfo{person}{Holger~R. Roth}, \bibinfo{person}{Shadi Albarqouni}, \bibinfo{person}{Spyridon Bakas}, \bibinfo{person}{Mathieu~N. Galtier}, \bibinfo{person}{Bennett~A. Landman}, \bibinfo{person}{Klaus {Maier-Hein}}, \bibinfo{person}{S{\'e}bastien Ourselin}, \bibinfo{person}{Micah Sheller}, \bibinfo{person}{Ronald~M. Summers}, \bibinfo{person}{Andrew Trask}, \bibinfo{person}{Daguang Xu}, \bibinfo{person}{Maximilian Baust}, {and} \bibinfo{person}{M.~Jorge Cardoso}.} \bibinfo{year}{2020}\natexlab{}.
\newblock \showarticletitle{The Future of Digital Health with Federated Learning}.
\newblock \bibinfo{journal}{\emph{npj Digital Medicine}} \bibinfo{volume}{3}, \bibinfo{number}{1} (\bibinfo{date}{Sept.} \bibinfo{year}{2020}), \bibinfo{pages}{1--7}.
\newblock
\showISSN{2398-6352}


\bibitem[S{\'a}nchez et~al\mbox{.}(2012)]%
        {sanchez2012playability}
\bibfield{author}{\bibinfo{person}{Jos{\'e} Luis~Gonz{\'a}lez S{\'a}nchez}, \bibinfo{person}{Francisco Luis~Guti{\'e}rrez Vela}, \bibinfo{person}{Francisco~Montero Simarro}, {and} \bibinfo{person}{Natalia Padilla-Zea}.} \bibinfo{year}{2012}\natexlab{}.
\newblock \showarticletitle{Playability: analysing user experience in video games}.
\newblock \bibinfo{journal}{\emph{Behaviour \& Information Technology}} \bibinfo{volume}{31}, \bibinfo{number}{10} (\bibinfo{year}{2012}), \bibinfo{pages}{1033--1054}.
\newblock


\bibitem[Silva et~al\mbox{.}(2023)]%
        {silva2023user}
\bibfield{author}{\bibinfo{person}{Nicollas Silva}, \bibinfo{person}{Thiago Silva}, \bibinfo{person}{Heitor Werneck}, \bibinfo{person}{Leonardo Rocha}, {and} \bibinfo{person}{Adriano Pereira}.} \bibinfo{year}{2023}\natexlab{}.
\newblock \showarticletitle{User cold-start problem in multi-armed bandits: When the first recommendations guide the user’s experience}.
\newblock \bibinfo{journal}{\emph{ACM Transactions on Recommender Systems}} \bibinfo{volume}{1}, \bibinfo{number}{1} (\bibinfo{year}{2023}), \bibinfo{pages}{1--24}.
\newblock


\bibitem[Soutter and Hitchens(2016)]%
        {soutter2016relationship}
\bibfield{author}{\bibinfo{person}{Alistair Raymond~Bryce Soutter} {and} \bibinfo{person}{Michael Hitchens}.} \bibinfo{year}{2016}\natexlab{}.
\newblock \showarticletitle{The relationship between character identification and flow state within video games}.
\newblock \bibinfo{journal}{\emph{Computers in human behavior}}  \bibinfo{volume}{55} (\bibinfo{year}{2016}), \bibinfo{pages}{1030--1038}.
\newblock


\bibitem[Stoffel et~al\mbox{.}(2022)]%
        {stoffelFederatedLearningHealthcare2022}
\bibfield{author}{\bibinfo{person}{AntunesRodolfo Stoffel}, \bibinfo{person}{Andr{\'e} da CostaCristiano}, \bibinfo{person}{K{\"u}derleArne}, \bibinfo{person}{YariImrana Abdullahi}, {and} \bibinfo{person}{EskofierBj{\"o}rn}.} \bibinfo{year}{2022}\natexlab{}.
\newblock \showarticletitle{Federated {{Learning}} for {{Healthcare}}: {{Systematic Review}} and {{Architecture Proposal}}}.
\newblock \bibinfo{journal}{\emph{ACM Transactions on Intelligent Systems and Technology (TIST)}} (\bibinfo{date}{May} \bibinfo{year}{2022}).
\newblock


\bibitem[Tompkins and Martins(2022)]%
        {tompkins2022masculine}
\bibfield{author}{\bibinfo{person}{Jessica~E Tompkins} {and} \bibinfo{person}{Nicole Martins}.} \bibinfo{year}{2022}\natexlab{}.
\newblock \showarticletitle{Masculine pleasures as normalized practices: Character design in the video game industry}.
\newblock \bibinfo{journal}{\emph{Games and Culture}} \bibinfo{volume}{17}, \bibinfo{number}{3} (\bibinfo{year}{2022}), \bibinfo{pages}{399--420}.
\newblock


\bibitem[Venkatasubramanian et~al\mbox{.}(2023)]%
        {venkatasubramanianIoTMalwareAnalysis2023}
\bibfield{author}{\bibinfo{person}{Madumitha Venkatasubramanian}, \bibinfo{person}{Arash~Habibi Lashkari}, {and} \bibinfo{person}{Saqib Hakak}.} \bibinfo{year}{2023}\natexlab{}.
\newblock \showarticletitle{{{IoT Malware Analysis Using Federated Learning}}: {{A Comprehensive Survey}}}.
\newblock \bibinfo{journal}{\emph{IEEE Access}}  \bibinfo{volume}{11} (\bibinfo{year}{2023}), \bibinfo{pages}{5004--5018}.
\newblock
\showISSN{2169-3536}


\bibitem[Wang et~al\mbox{.}(2020)]%
        {wangFederatedLearningMatched2020}
\bibfield{author}{\bibinfo{person}{Hongyi Wang}, \bibinfo{person}{Mikhail Yurochkin}, \bibinfo{person}{Yuekai Sun}, \bibinfo{person}{Dimitris Papailiopoulos}, {and} \bibinfo{person}{Yasaman Khazaeni}.} \bibinfo{year}{2020}\natexlab{}.
\newblock \bibinfo{title}{Federated {{Learning}} with {{Matched Averaging}}}.
\newblock
\newblock
\showeprint[arxiv]{2002.06440}~[cs, stat]


\bibitem[Wang et~al\mbox{.}(2023)]%
        {wangEffect2023}
\bibfield{author}{\bibinfo{person}{Jialin Wang}, \bibinfo{person}{Rongkai Shi}, \bibinfo{person}{Wenxuan Zheng}, \bibinfo{person}{Weijie Xie}, \bibinfo{person}{Dominic Kao}, {and} \bibinfo{person}{Hai-Ning Liang}.} \bibinfo{year}{2023}\natexlab{}.
\newblock \showarticletitle{Effect of Frame Rate on User Experience, Performance, and Simulator Sickness in Virtual Reality}.
\newblock \bibinfo{journal}{\emph{IEEE Transactions on Visualization and Computer Graphics}} \bibinfo{volume}{29}, \bibinfo{number}{5} (\bibinfo{year}{2023}), \bibinfo{pages}{2478--2488}.
\newblock


\bibitem[Watson(2020)]%
        {watsonDeepLearningTechniques2020}
\bibfield{author}{\bibinfo{person}{Alexander Watson}.} \bibinfo{year}{2020}\natexlab{}.
\newblock \bibinfo{title}{Deep {{Learning Techniques}} for {{Super-Resolution}} in {{Video Games}}}.
\newblock
\newblock
\showeprint{2012.09810}~[cs, eess]


\bibitem[Zhang et~al\mbox{.}(2023)]%
        {zhangFederatedFeatureSelection2023}
\bibfield{author}{\bibinfo{person}{Xunzheng Zhang}, \bibinfo{person}{Alex Mavromatis}, \bibinfo{person}{Antonis Vafeas}, \bibinfo{person}{Reza Nejabati}, {and} \bibinfo{person}{Dimitra Simeonidou}.} \bibinfo{year}{2023}\natexlab{}.
\newblock \showarticletitle{Federated {{Feature Selection}} for {{Horizontal Federated Learning}} in {{IoT Networks}}}.
\newblock \bibinfo{journal}{\emph{IEEE Internet of Things Journal}} \bibinfo{volume}{10}, \bibinfo{number}{11} (\bibinfo{date}{June} \bibinfo{year}{2023}), \bibinfo{pages}{10095--10112}.
\newblock
\showISSN{2327-4662}


\end{thebibliography}
